%% file: DOE_JSV.tex
\documentclass[preprint,12pt]{elsarticle}

\usepackage{graphicx}
\usepackage{amssymb}
\usepackage{multirow}
\usepackage{array}
\usepackage{subfig}
\usepackage{subfiles} 
\usepackage{tikz}
\usepackage{amsmath}
\usepackage{textcomp}

\definecolor{yellow}{rgb}{0.75, 0.75, 0.0}

\journal{Journal of Sound and Vibration}
\begin{document}

\begin{frontmatter}


\title{Effect of different geometries of self-oscillating trailing-edge flaplets on aerofoil self-noise}
\author[label1]{Edward Talboys}
\author[label2]{Thomas F. Geyer} 
\author[label2]{Florian Pr\"{u}fer}
\author[label1]{Christoph Br\"{u}cker}
\address[label1]{City, University of London, London, UK}
\address[label2]{Brandenburg University of Technology, Cottbus, DE}

\begin{abstract}
This paper presents an acoustic study of a standard NACA 0012 aerofoil with additional self-oscillating passive flaplets attached to the trailing edge. The tests with varying geometries of the flaplets were conducted in the anechoic wind tunnel at Brandenburg University of Technology, at chord based Reynolds numbers, $Re_c = 100,000 - 900,000$ at three geometric angles of attack $\alpha_g = 0^\circ, 10^\circ$ and $15^\circ$.
It was observed that all flaplet configurations reduce tonal noise and that the key geometric parameter to reduce this noise component is the width of the flaplets. 
The narrowest configuration tested almost completely removed the tonal noise, leading to an average overall sound pressure level reduction of up to 9~dB across the entire $Re_c$ range at $\alpha_g = 10^\circ$.
It was also observed that, in the low frequency regime, a further noise reduction  can be achieved by tuning the natural frequency of the oscillating flaplets. The thereby affected frequency range in the noise spectrum moves to higher frequencies when the natural frequency of the flaplets is increased and vice versa. 
Hence we show a novel way to target specific frequencies in passive aerofoil self-noise cancellation.  
\end{abstract}

\begin{keyword}
Aeroacoustics \sep Bioinspiration \sep Self-oscillating flaplets \sep Aerofoil self-noise


\end{keyword}

\end{frontmatter}


\section{Introduction}
Engineers have been extensively researching ways to mitigate aerofoil self-noise in recent years. 
There are various different sources of aerofoil self-noise, as explained in detail by \citet{Brooks1989}, but the main source of noise is boundary layer -- trailing edge interaction. 
For the current study, laminar boundary layer -- trailing edge noise is present and this manifests itself as a strong tonal noise, which is particularly annoying for the human hearing spectrum. 
As such this tonal noise has had a significant amount of research, to obtain a better understanding of this phenomenon and how it can be avoided. 

The first study to investigate in detail tonal noise was carried out by \citet{Paterson1972}.
They observed that there was an interesting feature, which they named `laddering', occurring. 
The main tonal peak frequency was scaling with the freestream velocity, $f \propto U_\infty^{0.8}$, until a certain point when the tonal noise peak made a sudden jump to a high frequency, hence termed laddering.
After averaging out all of these laddering events, across a wide range of velocities and frequencies, it was seen that $f \propto U_\infty^{1.5}$.
\citet{Tam1974} then proposed that the tonal noise is due to an acoustic feedback loop from an aeroacoustic interaction between instabilities in the boundary layer and noise sources situated in the aerofoil wake. 
The feedback model was expanded by \citet{ARBEY1983}, who showed that Tollmien-Schlichting (T-S) waves in the boundary layer diffract at the trailing edge of the aerofoil, subsequently creating acoustic waves that back-scatter upstream, feeding back into the feedback loop. 
This conclusion initiated more detailed investigations into the flow field around the aerofoil. 
\citet{Lowson} and \citet{McAlpine1999} first showed that a necessary feature of tonal noise is the presence of a laminar separation bubble on the pressure side of the aerofoil. They then further found that by using linear stability theory, the frequency of the tonal noise component was that of the most amplified instability within the boundary layer. 
\citet{Desquesnes2007} carried out the first direct numerical simulation (DNS) on the tonal noise issue. 
It was found that there was another, co-existing, feedback loop coming from the instabilities in the boundary layer on the suction side of the aerofoil. 
It was then thought that this feedback loop modulated the fringe frequencies that are found on either side of the tonal peak. 
\citet{Probsting2015} carried out simultaneous Particle Image Velocimetry (PIV) and acoustic measurements to find that at very low chord based Reynolds numbers (Re$_{\text{c}}$ = 30,000) the tonal noise is generated by the suction side of the aerofoil, while the pressure side dominates the generation of the tonal noise at higher Reynolds numbers (Re$_{\text{c}}$ = 230,000).
By tripping either side of the aerofoil separately, it is demonstrated that both feedback loops can exist independently \citep{Desquesnes2007}.  
\citet{Arcondoulis2019} built upon this dual feedback loop hypothesis to present an updated feedback model, where the tonal frequency is generated on both sides of the aerofoil.
It was observed that these tones have a near exact frequency to each other, but the fringe frequencies were seen to have large differences when comparing either side of the aerofoil. 
Their experimental observations were then used to develop a new dual feedback prediction model. 

Most of the boundary layer -- trailing edge mitigation strategies applied by researchers are inspired from the well known `silent' owl flight.  
One technique the owl uses, is due to it's `soft downy feather'. 
These features can be seen as acting like a porous surface, and Geyer et al. \cite{Geyer2010, Geyer2019} investigated the aeroacoustic benefit of having a porous aerofoil. 
It was found that by even having a small amount of porosity, either by a small streamwise amount of porosity \citep{Geyer2019} or with a high flow resistivity (low porosity) \citep{Geyer2010}, a benefit can be seen in the low -- mid frequency range. 
As the porosity increases, the benefit increases reaching up 10~dB broadband noise reduction. 
The benefits do come at a penalty, where there is an increase in high frequency noise.
This is primarily due to the increased surface roughness that the aerofoil is then constructed of. 

Another owl-inspired technique uses trailing edge brushes or serrations, mimicking the characteristic trailing edge structure formed by the feathers of owls. 
\citet{Herr2007}, for which an acoustic reduction was observed in the high frequency range (2--16~kHz).
This is believed to be due to the broadband noise of the turbulent boundary layer trailing edge interaction being affected. 
\citet{Finez2010} could show that the spanwise coherence of the shed vorticies in the wake behind the trailing edge is reduced by 25\% in the presence of brushes. 

Serrations have been extensively researched in both the laminar boundary layer case \citep{Chong2010, Chong2013b} and turbulent boundary layer case \citep{Leon2017, Leon2017b}.
For the laminar case, Chong et al. \cite{Chong2010, Chong2013b} found that the flat plate extensions were effective at reducing tonal noise by modifying the separation bubble on the pressure side of the NACA 0012 used. Another type of serration that has been seldom studied is slit-serrations. 
\citet{Gruber2010} carried out a comparative experiment on a tripped cambered NACA651210 aerofoil with saw-tooth serrations and slit serrations.
They found that reductions of up to 3~dB can be achieved with the slit serrations at low frequencies but increase frequencies at high frequencies.
It was seen that the maximum reduction occurs as the thickness is reduced and the spacing between the slits needs to be small for optimal noise reduction.

Studies with a single flexible flap at the trailing edge were investigated numerically by \citet{Schlanderer2013}. 
They carried out a DNS study on a flat plate with an elastic compliant trailing edge and found an aeroacoustic benefit at low and medium frequencies with an increased noise level at the Eigen frequency of the material. 
These results were confirmed later by \citet{Das2015} in an experimental investigation using a similar arrangement to \citet{Schlanderer2013}.
In a recent study from \citet{Yu2020}, a 2D simulation on an S833 aerofoil with an oscillating trailing edge fringe was carried out.
By controlling the Eigen frequency of the flapping it was shown that the strength and size of the shed vorticies was reduced.
They also showed that the lift coefficient was increased and drag coefficient was reduced, and this was attributed to the creation of a suction region downstream of the fringe.  
In a similar manner, active oscillations of a trailing edge flap were studied by \citet{Simiriotis2019}. 
Their investigation was focused on the wake structure and it was observed that the wake could be reduced in thickness by as much as 10\%.

In the present study, an array of individual self-oscillating elastic elements are attached to the trailing edge of a symmetric NACA~0012 airfoil, in an attempt to mimic the tips of birds feathers. 
This type of trailing edge modification with arrays of individual mechanical oscillators in form of elastic flaps has thus far only studied by the authors \citep{Kamps2017a,Kamps2017, Geyer2019a, Talboys2018, Talboys2019, Talboys2019a}. 
Benefits from tonal noise reduction with these type of oscillators has been previously seen \citep{Kamps2017, Talboys2019} and was attributed to a dampening of T-S instabilities within the boundary. 
Using high-speed PIV measurements in the boundary layer along the suction side of the aerofoil,
\citet{Talboys2018} showed that non-linear instabilities within the shear layer were suppressed by a lock-in process, in which the natural frequency of the oscillating flaplets locks with the fundamental instability (flaplets act as pacemaker).
This ultimately led to a reduction in the boundary layer thickness, when the flaplets were applied, due to the reduced probability of paired vorticies in the boundary-layer.

The present investigation will provide further insight into the aeroacoustic effect of these flaplets, where the focus is the variation of the geometrical and material parameters. 
To this end, the length, width and inter-spacing of these flaplets are altered and the acoustic effect as well as the effect on the flow field, via hot wire anemometry, are detailed herein.

\section{Experimental Arrangement and Measurement Techniques}
\subfile{Fig_Tex/MicArray}
The experiments were carried out in an open jet style wind tunnel at Brandenburg University of Technology, Cottbus \cite{Sarradj2009}. Schematics and a photograph of the set-up are shown in Fig.~\ref{fig: Set-up}.
The wind tunnel was equipped with a circular nozzle with a contraction ratio of 16 and an exit diameter of 0.2~m. 
With this nozzle, the maximum flow speed is in the order of 90~m/s and at 50~m/s, the turbulence intensity in front of the nozzle is below 0.1~\%. 
For the present study the chord based Reynolds number ($Re_c$) was varied from 100,000 -- 900,000 and the geometric angle of attack, $\alpha_g$ , was varied from $\alpha_g = 0^\circ$ -- $15^\circ$.

The aerofoil used is a NACA~0012 symmetric aerofoil, with a chord of 0.19~m and a span of 0.28~m.
The span is such that it extends the entire nozzle diameter, to ensure that there are no wing tip effects.
The aerofoil was 3D printed, using a polyjet printer, in two halves, where the dividing line was the chord (centre) line of the aerofoil. 
This allowed the flaplets to be inserted easily along the chord line and extrude out of the trailing edge of the aerofoil, such that the free ends were orientated downstream (see Fig.~\ref{fig: Aerofoil Pic}). 
This allows them to freely oscillate at their Eigen frequency (equal to the natural frequency).
The determination of the Eigen frequency will be discussed in Section~\ref{sec:Results sub:Static}. 
The flexible trailing edge flaplets were manufactured, using a laser cutter, from a thin polyester film (see Table \ref{table: Flaplets Spec} for dimensions). 
It should be noted here that a singular flexible trailing edge, one that spans the entire trailing edge, was not tested in this study.
This was motivated by the idea to break-up any spanwise coherence. 

The acoustic measurements were performed using a planar microphone array consisting of 56 1/4 inch microphone capsules flush mounted into an aluminium plate with dimensions of 1.5~m~\texttimes~1.5~m. The array was positioned in a distance of about 0.7~m above the aerofoil (see Fig.~\ref{fig: Exp Arr. Acc}). Data from the 56 microphones were recorded with a sampling frequency of 51.2~kHz and a duration of 60~s using a National Instruments 24 Bit multi-channel measurement system.
To account for the refraction of sound at the wind tunnel shear layer, a correction method was applied that is based on ray tracing \cite{Sarradj2017}. 
In post processing, the time signals were transferred to the frequency domain using a Fast Fourier Transformation (Welch’s method, \cite{Welch1967}), which was done block-wise on Hanning-windowed blocks with a size of 16384 samples and 50\% overlap.
This lead to a small frequency spacing of only 3.125~Hz.
The resulting microphone auto spectra and cross spectra were averaged to yield the cross spectral matrix.
This matrix was further processed using the DAMAS beamforming algorithm proposed by \citet{Brooks2006}, which was applied to a two-dimensional focus grid parallel to the array and aligned with the trailing edge. 
The grid has a streamwise extent of 0.12~m, a spanwise extent of 0.12~m and an increment of 0.02~m. 
The increment was chosen as there was seen to be minimal differences between grids with an increment of 0.005~m and 0.01~m. 
Since the amount of experimental data collected in the present study is quite large, it was decided to use the coarser grid, which has great computational benefits as the calculation scales with the square of the size of the grid.
Saying this, a grid with 0.01~m increments was used for the 2D sound maps for better image resolution. 
The outcome of the beamforming algorithm is a two-dimensional map of noise source contributions from each grid point, a so-called sound map. 
In order to obtain spectra of the noise generated by the interaction of the boundary layer with the trailing edge of the aerofoil, a sector was defined that only contains the noise source of interest. 
The chosen sector has a chordwise extent of 0.2~m and a spanwise extent of 0.1~m. 
Thus, spectra of the noise generated by this mechanism are derived by integrating all noise contributions from within this sector, while all potential background noise sources (such as the wind tunnel nozzle or the aerofoil leading edge) are excluded from the integration. 
The resulting sound pressures were then converted to sound pressure levels, $L_p$ (with a reference value of 20~$\mu$Pa), and 6~dB were subtracted to account for the reflection at the rigid microphone array plate. 

The hot wire anemometry (HWA) measurements were taken in separate experiments to the acoustic measurements, to insure no additional noise from the HWA and associated traverse system was measured in the acoustic spectra. The probe used was a Dantec X wire probe (55P64), where the data was taken at a sampling frequency of 25.6~kHz. The Dantec HWA hardware system used for the measurements contains an electronic low-pass filter with a cut-off frequency of 10~kHz. The wake profiles were initiated at 0.25$c$ above the aerofoil till 0.25$c$ below the aerofoil, at a distance of 0.25$c$ from the solid aerofoil edge approximately at mid span. The increment of the measurements was large in the freestream region (1~mm), and then the increment was systematically reduced such that the region of the boundary layer was measured with 0.2~mm increments. Each measurement was taken for a period of 10~s, prior to moving on to the next increment. 

\begin{table}[t]
\small
\centering
\begin{tabular}{lccc}
\hline
\multirow{2}{*}{\textbf{Name}} & \textbf{Length (L)} & \textbf{Width (w)} & \textbf{Spacing (d)} \\
                               & \textbf{[mm]}   & \textbf{[mm]}  & \textbf{[mm]}      \\ \hline
Baseline                       & 20              & 5              & 1 \\
Long                           & 30              & 5              & 1 \\
Short                          & 10              & 5              & 1 \\
Wide                           & 20              & 10             & 1 \\
Narrow                         & 20              & 2.5            & 1 \\
Large Spacing                  & 20              & 5              & 7 \\
Medium Spacing                & 20              & 5              & 3 \\ \hline
\end{tabular}
\caption{Geometric specifications and naming convention for the tested flaplet cases. The thickness of each flaplet configuration was constant at 0.18~mm.}\label{table: Flaplets Spec}
\end{table}






\section{Results}\label{sec:Results}
\subsection{Static Flaplet Response}\label{sec:Results sub:Static}
\begin{figure}[t] 
\centering\includegraphics[scale = 1]{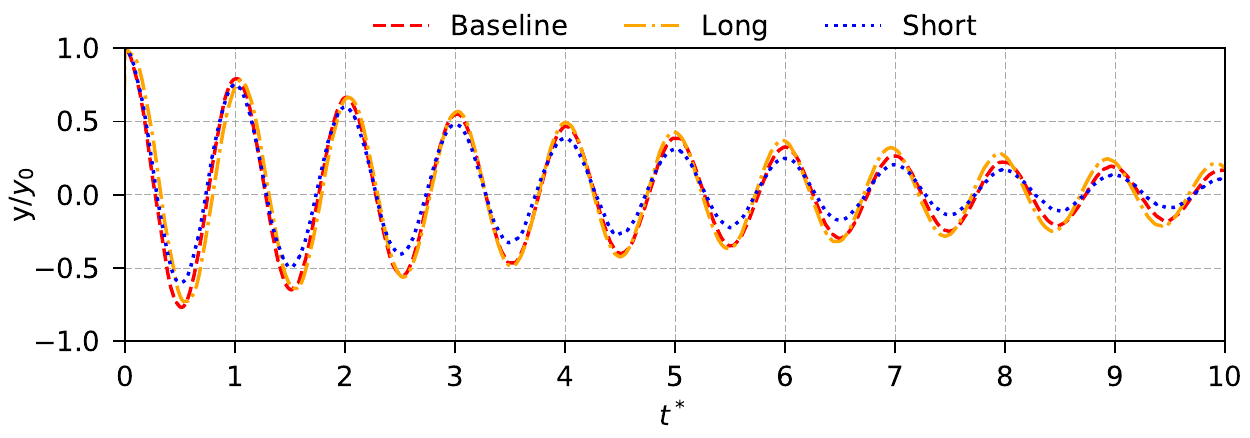}
\caption{Response of the flaplets to a step input, normalised with the stating position and the measured Eigen frequency.}\label{fig: Static}
\end{figure}

The Eigen frequency of the flaplets needs to be determined in order to see at what frequency the flaplets will oscillate in the flow. 
The set-up and image processing for this response test is the same as was used in \cite{Talboys2018}.
The normalised response of the long, baseline and short flaplets is shown in Fig.~\ref{fig: Static}, where $t^* = t \cdot\omega_n$, t is the time in seconds, $\omega_n$ is the observed Eigen frequency in Hz, $y$ is the displacement and $y_0$ is the starting position of the flaplet. 
Once the Eigen frequency is obtained, the Young's Modulus of the flaplets can be derived using classical cantilever beam theory. 
Determining the Young's modulus then allows the use of a finite element analysis (FEA) software, in the present case ANSYS, in order to predict the first torsion mode. 
This is when the flap response enters the flutter regime and therefore provides a limit of frequencies, above which any excitation of this mode may disturb the lock-in effect and add to additional noise.
Therefore, the range of tested flow speeds (Reynolds-number) was limited to avoid the excitation of this mode.

\begin{table}[t]
\small
\begin{tabular}{lccc}
\hline
\multirow{2}{*}{Case} & Experimental             & \multicolumn{2}{c}{Finite Element Analysis}   \\
                      & Eigen Freq. [Hz] & Bending Mode [Hz] & Torsion Mode [Hz] \\ \hline
Baseline              & 107                  & 110                   & 918                   \\
Long                  & 50                   & 45                    & 547                   \\
Short                 & 345                  & 369                   & 1702                  \\
Wide                  & 105                  & 111                   & 494                   \\
Narrow                & 128                  & 118                   & 1879                  \\
Large Spacing           & 104                  & 99                    & 919                   \\
Medium Spacing            & 111                  & 110                   & 828                   \\ \hline
\end{tabular}
\caption{Experimental Eigen frequencies and finite element analysis predictions of both the 1$^{\text{st}}$ bending and 1$^{\text{st}}$ torsion modes.}\label{table: Static}
\end{table}

From Table \ref{table: Static}, each of the flaplets Eigen frequencies can be seen, derived both experimentally and by using FEA. 
It can be observed that there is good agreement between the FEA model and the experiment. 
As the Eigen frequency is inversely proportional to the squared length, only the short and long flaplets should differ largely in their Eigen frequency. 
It is important to note that the narrow cases show a higher Eigen frequency than expected. This is assumed to be caused by the laser-cutting process, when additional heat is introduced at the cutting edge. 
As a result, the Young's modulus of the narrow flaplets was seen to be slightly different to the other flaplets and was adjusted accordingly in the FEA model. 
When looking at the torsional mode, it can be seen that the long and the wide flaplet cases have a considerably lower torsional frequency in comparison to the other flaplets.
It can therefore be expected that those will go into flutter at a lower Reynolds number, hence limiting the working velocity range for these cases.  

\subsection{Tonal Noise}
\label{sec:Results sub:Tonal}
\begin{figure}[t!] 
\centering\includegraphics[scale = 1]{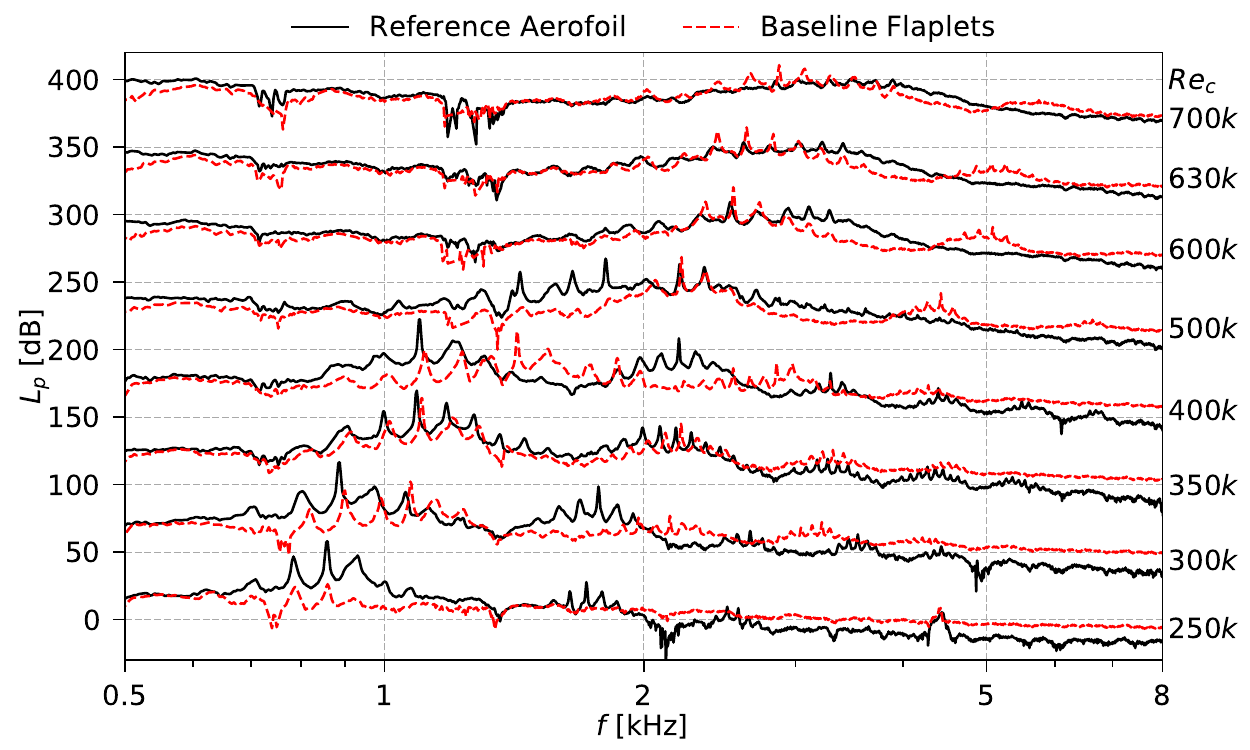}
\caption{Far field narrow-band sound pressure level spectra $re~20\mu$Pa for the reference (plain) aerofoil, focusing on the tonal noise produced at $\alpha_g=10^\circ$. Each of the spectra are spaced with 50~dB from each other for clarity.}\label{fig: Narrowband-Baseline}
\end{figure}

\begin{figure}[t!] 
\centering\includegraphics[scale = 1]{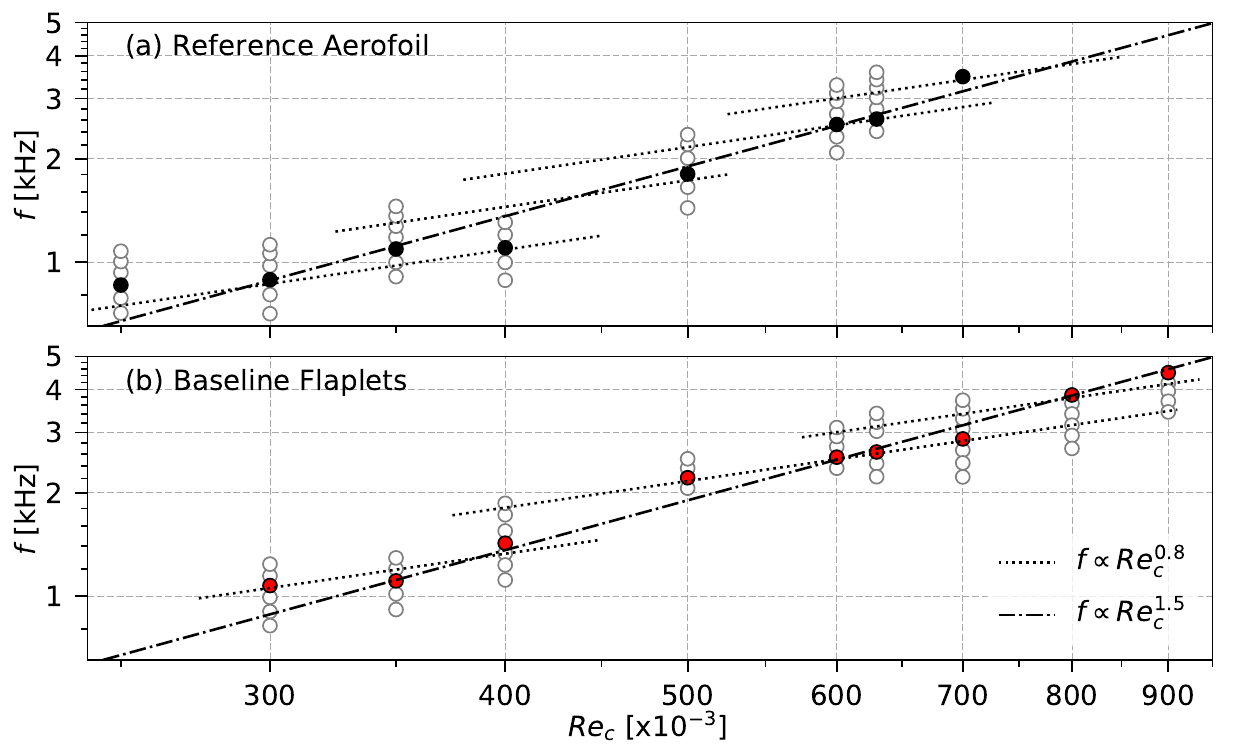}
\caption{Tonal peaks (solid) and fringe frequencies (hollow) for each Reynolds number at $\alpha_g = 10^\circ$. Global and local trend lines from \citet{Paterson1972} are also indicated.}\label{fig: Tonal Peaks}
\end{figure}

For the current study tonal noise was seen at all three geometric angles of attack, but for brevity only the results for the reference aerofoil and the aerofoil with baseline flaplets at $\alpha_g=10^\circ$ are shown here. 
Figure~\ref{fig: Narrowband-Baseline} shows the narrow-band far field acoustic spectra for the chord based Reynolds number where tonal noise was observed, $250,000 \leq Re_c \leq 700,000$. 
Each of the spectra are spaced with 50~dB from each other for clarity. 
One of the key features that is observed with tonal noise is distinct frequencies overlaid on a broadband hump. 
These distinct peaks can be seen in all of the presented cases in Fig.~\ref{fig: Narrowband-Baseline}.
As the $Re_c$ is increased, the range of these peaks has moved to a higher frequency, too.
Another feature that is well observed in tonal noise is the `laddering' effect.
This can be clearly seen in the spectra obtained for the reference aerofoil, where the frequency of the tonal peaks from $Re_c = 250,000$ -- $400,000$ increases linearly. 
To further analyse this effect, Fig.~\ref{fig: Tonal Peaks} shows the frequency of the tonal peak and the associated fringe frequencies at each $Re_c$.
The plot includes trend lines both for the local velocity-frequency scaling and for the global velocity-frequency scaling, as proposed by \citet{Paterson1972}, revealing the typical `rungs' of the ladder for both the reference and baseline case.

When comparing the reference aerofoil and baseline flaplets tonal noise, it can be seen that the flaplets indeed reduce the magnitude of the tonal noise component, consistent for the whole range of tested Reynolds numbers. 
In the lowest $Re_c $ case presented, $Re_c = 250,000$, the tonal noise has been delayed to a higher Reynolds number with the addition of the flaplets. Furthermore, the tonal peaks extend to a higher velocity than the reference aerofoil due to the presence of the flaplet. In addition, it is visible that the spectra obtained for the reference aerofoil contain a second range of tones at twice the frequency of the first occurrence, which are the first harmonics of the tonal noise. These harmonics are much less distinct in the spectra obtained for the aerofoil with flaplets, especially at low $Re_c$.
These findings strongly indicate that flaplets are dampening the T-S instabilities within the boundary layer.

In order to enable a better description of the effect of the flaplets on the noise generation compared to the reference aerofoil, far-field noise spectra will be shown in 1/3 octave bands in the remainder of this paper.


\subsection{Acoustic and Hot Wire Anemometry Results}
In this section the results of the 1/3 octave band acoustic spectra, 2D sound maps, overall sound pressure level and hot wire wake measurements will be presented in this order, in turn, for each of the geometric variations, starting with the variation in length, then width and finally the inter-spacing.
\subsubsection{Variation in Flaplet Length}\label{sec:Results sub:Length}
\begin{figure}[t!] 
\centering\includegraphics[scale = 1]{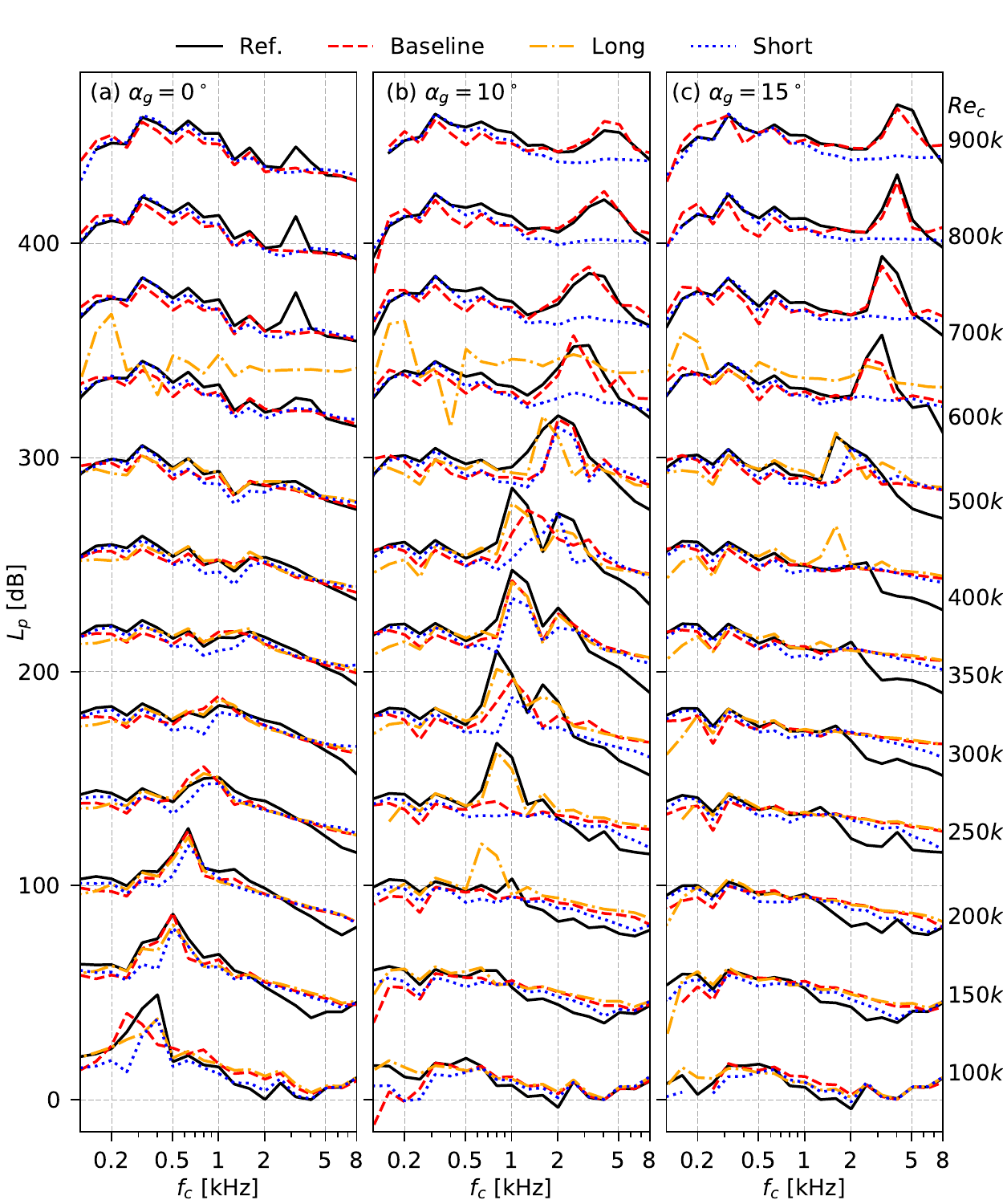}
\caption{1/3 Octave band acoustic spectra for variation in length. Each of the spectra are spaced with 35~dB from each other for clarity.}\label{fig: AS-Length}
\end{figure}

Figure \ref{fig: AS-Length} shows the acoustic 1/3 octave band spectra at three geometric angles of attack at various chord based Reynolds numbers. 
In order to visualise all Reynolds number cases on one plot, each spectra are incremented 35~dB from the spectrum at the previous Reynolds number. 
At $\alpha_g=0^\circ$, Fig.~\ref{fig: AS-Length}a, a clear tonal noise component is visible for all cases up to a Reynolds number of 250,000.
It can be seen that the baseline and long flaplets have little effect on the tonal noise at this angle of attack, whereas the short flaplets show a clear reduction in tonal noise.

What is interesting here is that there is seemingly a strong correlation of  Eigen frequencies of the flaplets and the frequency range of most efficient noise reduction. 
In Fig.~\ref{fig: AS-Length}, for the lowest frequencies investigated (100--200~Hz), the longest flaplets (lowest Eigen frequency) lead to the highest reduction. 
Then, at slightly higher frequencies around 00--500~Hz, the baseline flaplets show the most reduction, followed by the shortest flaplets (highest Eigen frequency) having the most reduction again at slightly higher frequencies (500--2000~Hz). 
This tendency clearly supports the hypothesis of the lock-in effect documented in \cite{Talboys2018} (flaplets act as pacemaker), which plays an important role in the stabilization of the T-S waves. Because of the different Eigen frequencies of the flaplets, they are modifying the wake with different frequencies, hence the different acoustic frequency reductions.
This type of vortex shedding noise reduction has also been observed by \citet{Geyer2019a}, where they used a cylinder with flexible flaplets on the aft part of the cylinder. 

An interesting feature can be seen at Reynolds number of 600,000, where the long flaplets show a dramatic increase in the noise level.
This is when this specific geometry starts to `flutter'. 
The flutter is due to the excitation of the torsional natural frequency of these flaplets, which is at a much lower frequency for the long flaplets in comparison to the others, see Table~\ref{table: Static}. 
This can be observed at the same Reynolds number for each of the geometric angles of attack, showing that this is indeed the critical Reynolds number, where the torsional mode is excited for this geometry. 
It can also be seen that all of the flaplets increase the noise level at frequencies above 5~kHz. 
Again, as at low frequencies, there is a distinct order in which the noise level is increased.
The most elevated levels are observed for the long flaplets, while the least comes from the short ones. 
At a Reynolds number of 700,000, there is an emergence of trailing-edge bluntness noise at approximately 3~kHz on the reference aerofoil.
As is to be expected, the flaplets reduce/remove this noise component, as they are effectively reducing the bluntness from $\approx$ 1~mm to the thickness of the flaplets (0.19~mm). 

As the angle is increased to $\alpha_g = 10^\circ$ (Fig.~\ref{fig: AS-Length}b), the emergence of tonal noise can be seen at a Reynolds number of 250,000 for the baseline case, and is present up to the highest Reynolds number tested (900,000), as previously discussed in Section~\ref{sec:Results sub:Tonal}. 
A similar trend as for the zero degree angle case is oberved: All the flaplets show a reduction in the low frequency range and an increase at the higher frequency range, the magnitude of this modification being tendentiously higher.

The tonal noise component is in general reduced at all flaplet cases, albeit that the long flaplets only show a small reduction in comparison to the reference. 
Saying this, at $Re_c$ = 600,000 and above the baseline flaplets are approximately at parity with the reference case and, interestingly, the tonal noise is completely removed for the short flaplets. 
The longest flaplets are seen to enable the initiation of tonal noise even before the reference case. 
This is believed to be due to the flaplets, at this Eigen frequency, promoting the formation of the separation on the pressure side of the aerofoil, the necessary mechanism for tonal noise.  

At the highest tested angle, $\alpha_g = 15^\circ$, the tonal noise on the reference aerofoil starts to occur at higher Reynolds numbers.
This is consistent with previous literature \cite{Talboys2019}.
All the trends that have been observed at $\alpha_g = 10^\circ$ can also be seen at the increased angle.
Again, the magnitude of the effects are increased further. 

\begin{figure}[t!] 
\centering\includegraphics[scale = 1]{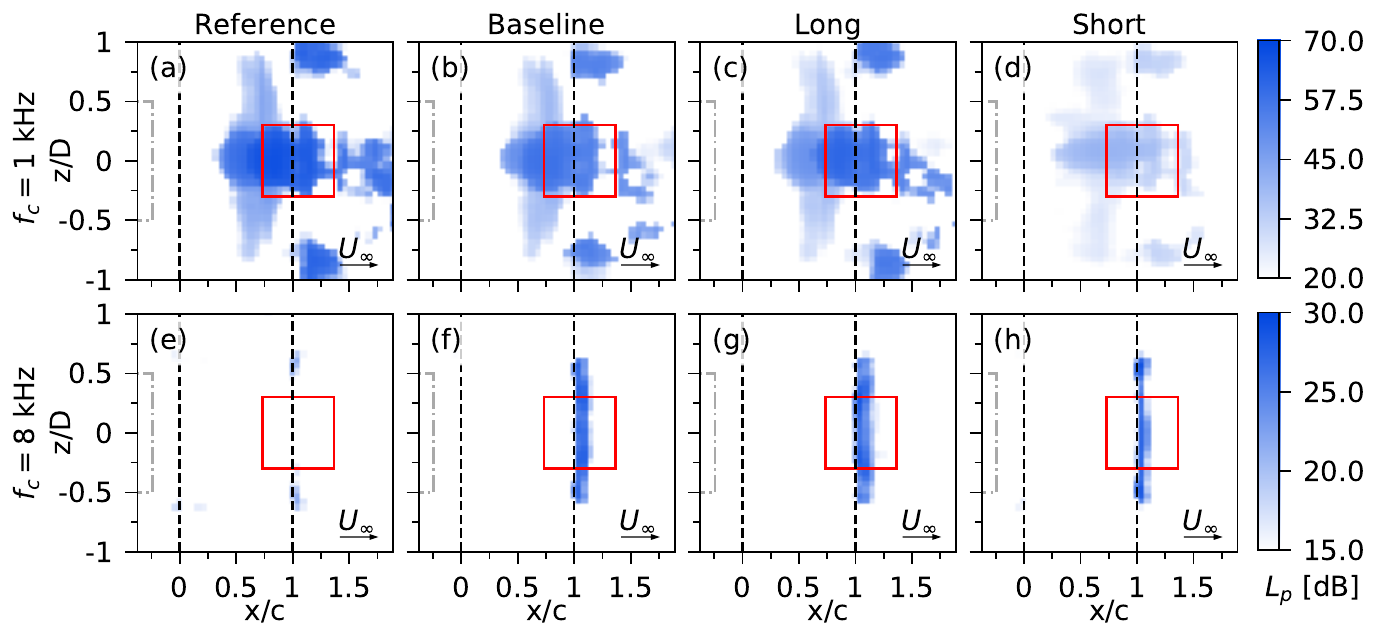}
\caption{2D sound maps as the length is varied, at $Re_c$=350,000 and $\alpha_g=10^\circ$. (a--d) shows the frequency band, $f_c$=1~kHz and (e--h) shows the frequency band, $f_c$ = 8~kHz. (\protect\begin{tikzpicture}
\protect\tikz[baseline=-2pt] \protect\draw[dash dot,very thick,color=lightgray] (0.0,0.1) -- (0.575,0.1);
\protect\end{tikzpicture}) indicates the jet nozzle, (\protect\begin{tikzpicture}
\protect\tikz[baseline=-2pt] \protect\draw[dashed,very thick,color=black] (0.0,0.1) -- (0.575,0.1);
\protect\end{tikzpicture}) indicates the aerofoil, (\protect\begin{tikzpicture}
\protect\tikz[baseline=-2pt] \protect\draw[solid,very thick,color=red] (0.0,0.1) -- (0.575,0.1);
\protect\end{tikzpicture}) indicates the interrogation region for acoustic spectra.}\label{fig: SoundMap-Length}
\end{figure}

In order to view the location of the dominant acoustic sources, sound maps can be used. 
These maps are shown as sound pressure levels on a discretised grid that can be interrogated to observe certain noise sources in certain frequencies bands. These sound maps are presented as plan views of the test set-up, and in each of the plots the nozzle exit, leading and trailing edge and the spectral integration zone are indicated.

Figure~\ref{fig: SoundMap-Length} shows such 2D sound maps obtained at $Re_c$ = 350,000 and $\alpha_g = 10^\circ$ for the frequency bands with center frequencies $f_c$ of 1~kHz and 8~kHz respectively. This specific testing condition has been chosen as a clear tonal peak can be seen in the acoustic spectra for the reference case, which therefore forms a good basis to compare any tonal noise reduction.

Tonal noise has been previously shown to be reduced with the presence of flaplets. 
From Fig.~\ref{fig: SoundMap-Length}a it can be seen that the noise source starts from the chordwise position x/c $\approx$ 0.3 extending beyond the solid trailing edge and is located in the mid-span location of the aerofoil.
It must be noted here that there are small regions of increased noise at either side of this large noise source.
These noise sources are due to the interaction of the shear layer of the wind tunnel, which is highly turbulent, with the trailing edge of the airfoil. 
Looking at all the flaplet cases it can be seen that with the baseline and long flaplets (Fig.~\ref{fig: SoundMap-Length}b-c) there is only a small reduction of the tonal noise, whereas, in contrast, the short flaplets (d) reduce the tonal noise significantly.

When investigating the locations of the acoustic sources with respect to the observed high frequency noise increase, the sound maps at a 1/3 octave band with $f_c$ = 8~kHz are better suited than those obtained at 1~kHz.
As seen in Fig.~\ref{fig: SoundMap-Length}e, at the higher frequency the reference case now only shows noise sources where the jet shear layer interacts with the leading edge and trailing edge of the aerofoil. 
It can then be seen for the flaplet cases that somewhat weaker noise sources appear directly aft of the solid trailing edge and along the length of the flaplets, which correspond to the increase in noise levels seen in the acoustic spectra in Fig.~\ref{fig: AS-Length}. 
This is thought to be due to the oscillatory motion of the flaplets giving rise to an additional acoustic source. 
It can be seen that this source is a function of the flaplets surface area, as there is a clear difference in the extension of the source from the trailing edge as the flaplets increase in length. 
Where (h), corresponding to the short flaplets, is the smallest region and the largest area is seen in (g) which is for the long flaplets.

\begin{figure}[t!] 
\centering\includegraphics[scale = 1]{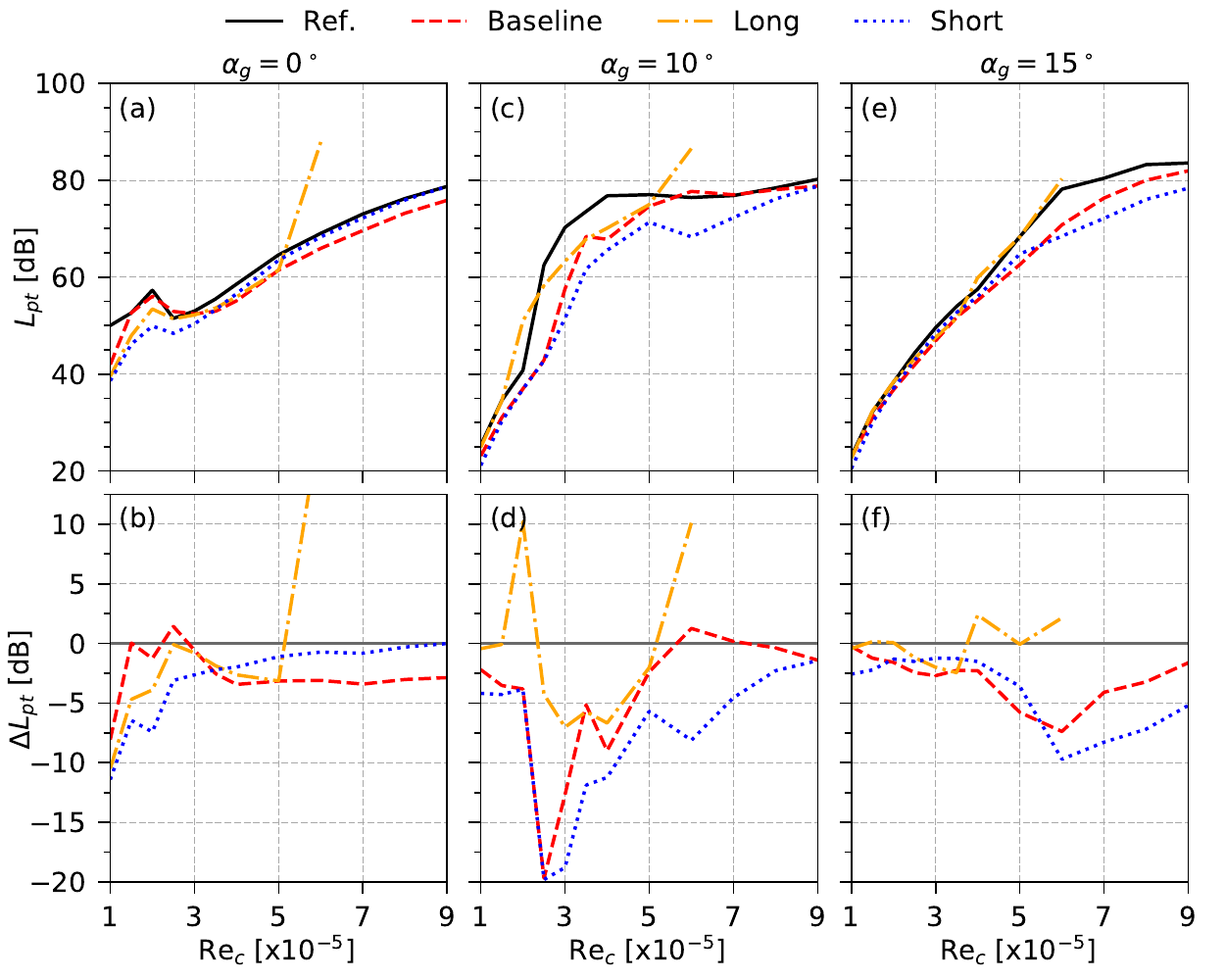}
\caption{OSPL, denoted as $L_{pt}$, and Delta OSPL, $\Delta L_{pt}$, for variation in length.}\label{fig: OSPL-Length}
\end{figure}
As a means to investigate the overall acoustic effect of the flaplets, the overall sound pressure level (OSPL) has been computed for each of the cases according to
\begin{equation}
L_{pt} = 10 \hspace{1mm} \log_{10} \Big(\sum_{f_c=0.2~\text{kHz}}^{8~\text{kHz}} 10^{\frac{L_{pi}}{10~\rm dB}}\Big)~~\text{dB}
\label{eqn: OSPL},
\end{equation}
where $L_{pi}$ is the sound pressure level at the $i^{th}$ centre frequency ($f_c$). In order to obtain an easier appreciation of the magnitude of the differences between the cases, the difference between the reference case and the different flaplets cases has also been calculated as
\begin{equation}
\Delta L_{pt} = L_{pt, \rm flaplet} - L_{pt, \rm reference}. \label{eqn: DOSPL}
\end{equation}
The results are shown in Figs.~\ref{fig: OSPL-Length}a-c-e. 

At $\alpha_g = 0^\circ$, Fig.~\ref{fig: OSPL-Length}a, it can be seen that as the Reynolds number increases, the general trend is that the OSPL also increases. 
There is a slight reduction in OSPL from $Re_c$= 200,000 -- 250,000, which is due to the tonal noise component disappearing at this condition. When comparing the flaplet cases to the reference case, Fig.~\ref{fig: OSPL-Length}b, it can be seen that at the lowest Reynolds numbers up to 350,000, the shortest flaplets show the best overall reduction.
This reduction is of the order of 5 to 10~dB. 
After this point the OSPL of the shortest flaps tends back to the reference case. 
The OSPL for the baseline flaplets, on the other hand, is initially approximately the same as that for the reference case and then tends to a reduction of $\sim$3.5~dB from a Reynolds number of 400,000 onward. 
The long flaplets show reasonable overall reductions at low Reynolds number until the sudden noise increase occurs at Reynolds numbers above 500,000, as can be seen in the acoustic spectra.
At $\alpha_g=10^\circ$, Fig.~\ref{fig: OSPL-Length}c, the reference case leads to a sudden increase in OSPL at $Re_c$= 250,000.
This is due to the prevalent tonal noise that dominates the spectra.
Once the Reynolds number reaches 400,000, the OSPL plateaus for the remaining tested Reynolds numbers. 
Immediately it can be seen that for the baseline and short flaplet cases, the OSPL is lower across the whole range of Reynolds numbers.
The largest reduction of approximately 20~dB can be seen at $Re_c$=250,000 and this is due to the delay in the occurrence of tonal noise, which can be clearly seen in the acoustic spectra (Fig. \ref{fig: AS-Length}b). 
At Reynolds numbers beyond 500,000 it is seen that the baseline flaplets do not show too much, if any, OSPL reduction, which is due to the tonal noise in both the reference and baseline cases being approximately the same.
As was seen at $\alpha_g=0^\circ$, the OSPL for the short flaplets tends towards the value for the reference aerofoil as the Reynolds number increases.
Increasing $\alpha_g$ further to 15$^\circ$ reveals similar effects as those seen for both lower angles, although the effects are less pronounced. 
At Reynolds numbers below 400,000 it can be seen that both the short and baseline flaplets show an OSPL reduction of around 2.5~dB, after which the reduction increases. 
This is again due to the lower tonal noise component observed with the flaplets.    

\begin{figure}[t!] 
\centering\includegraphics[scale = 1]{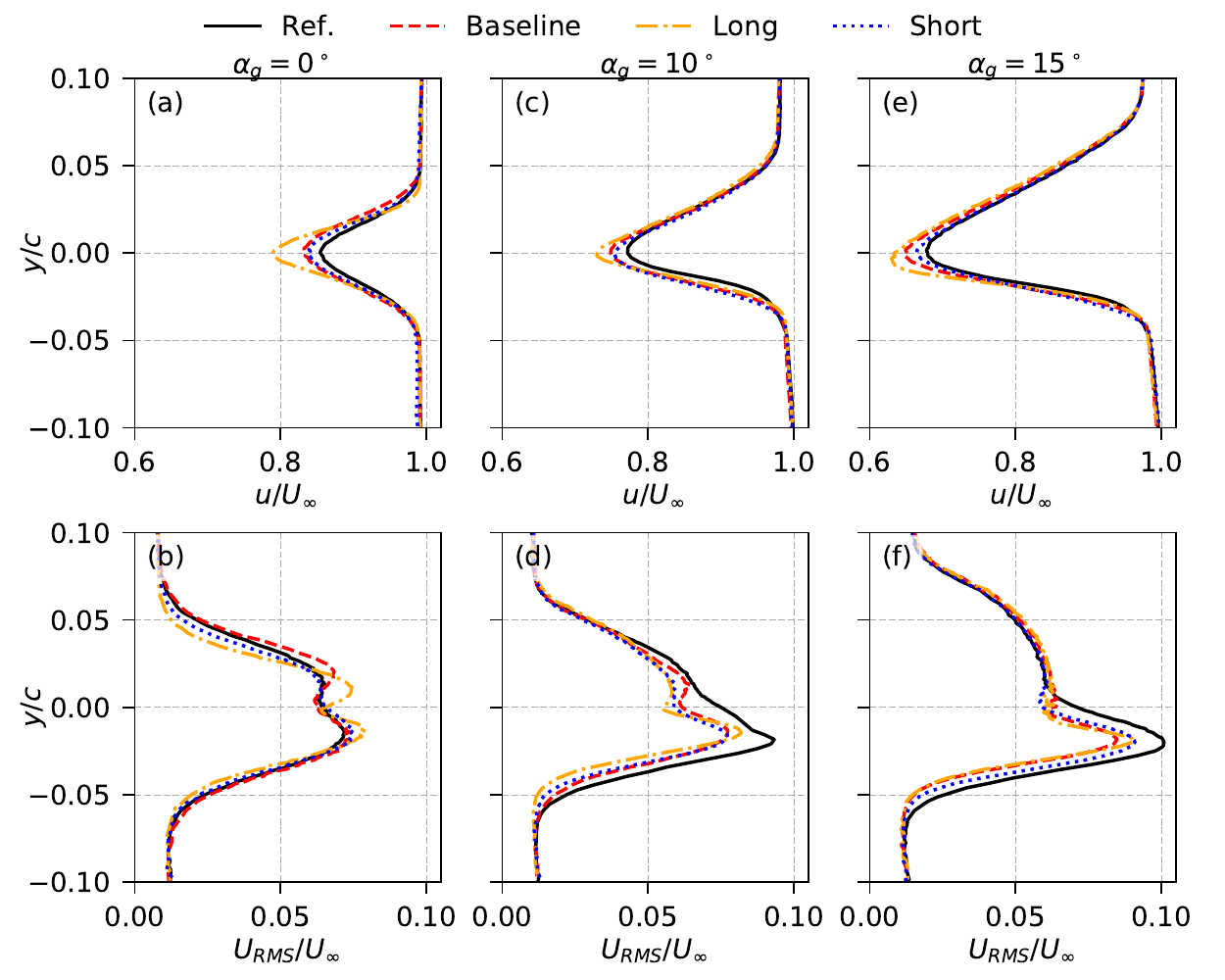}
\caption{Mean and RMS wake profiles of the streamwise velocity component at $Re_c = 200,000$ for variation in length.}\label{fig: HWA-Length}
\end{figure}

In order to observe the effect of the flaplets on the wake flow, hot wire measurements were taken at a streamwise distance of 0.25$c$ aft the solid trailing edge.
Therefore, no offset has been taken into account for the flaplets length.
Figures~\ref{fig: HWA-Length}a-c-e show mean streamwise velocity profiles for each of the flaplets of varied length at each of the tested angles of attack at $Re_c =200,000$.
At $\alpha_g = 0^\circ$, the profile is symmetric about the $y/c = 0$ line.
This is expected as the aerofoil is symmetric itself, the wake deficit should also be symmetric here.
As the angle is increased, the wake profiles become thicker on the suction side of the aerofoil ($y/c > 0$), which is due to the thickening of the boundary layer on the this side of the aerofoil. 
In addition, the absolute value of the velocity deficit is also seen to increase at higher angles, from a value of $u/U_\infty$~=~0.85 for the reference aerofoil at $\alpha_g = 0^\circ$ to a value of around 0.7 for the same aerofoil at $\alpha_g = 15^\circ$
It can be seen that at each of the angles the largest velocity deficit is visible for the long flaplets followed by the baseline and then the short flaplets.
This is a logical conclusion as there was no offset for the flaplet length, hence the tip of the long flaplet was physically closer to the probe. 
A more discernible difference between flaplets and reference aerofoil can be seen when looking at the streamwise RMS profiles as shown in Figs.~\ref{fig: HWA-Length}b-d-f.
At $\alpha_g=0^\circ$ it can be seen that all RMS profiles are similar, even with those obtained for the baseline and the long flaplets showing a small increase compared to the reference aerofoil.
As the angle increases to $\alpha_g = 10^\circ$, Fig.~\ref{fig: HWA-Length}d, a peak in the RMS velocities can be seen on the pressure side of the aerofoil. 
This is very evident for the reference aerofoil and is attributed to be due to the high levels of turbulence caused by the separation and reattachment of the separation bubble on the pressure side of the aerofoil close to the trailing edge.
All of the flaplet cases show some reduction in the RMS velocities and are of similar amplitude.
Increasing the angle further, again shows a similar level of reduction for each of the tested cases. 
Therefore it is clearly seen that the flaplets indeed modify the turbulent fluctuations in the wake at $\alpha_g=10^\circ$ and $15^\circ$.
Interestingly, there also does not seem to be a length dependency in this reduction.

\subsubsection{Variation in Flaplet Width}\label{sec:Results sub:Width}
\begin{figure}[t!] 
\centering\includegraphics[scale = 1]{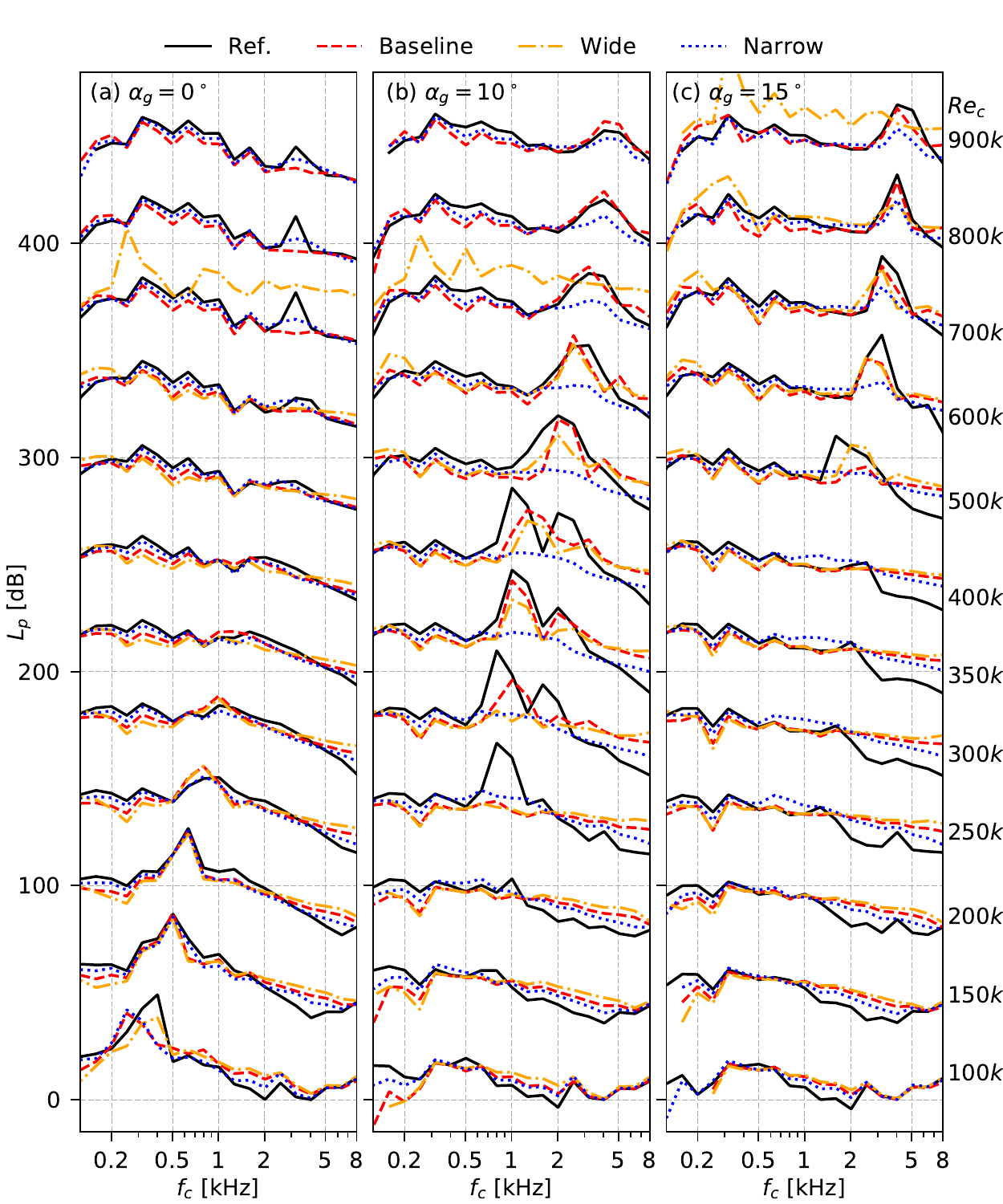}
\caption{1/3 Octave band acoustic spectra for variation in width. Each of the spectra are spaced with 35~dB from each other for clarity.}\label{fig: AS-Width}
\end{figure}

The next parameter varied was the spanwise width of the flaplets.
At $\alpha_g=0^\circ$, Fig.~\ref{fig: AS-Width}, it can be seen that for the cases up to $Re_c$=500,000 the low frequency reduction is most prominent when the wide flaplets are attached. 
This is followed by the baseline flaplet and then the narrow flaplets.
For higher Reynolds numbers, the wide flaplets sound pressure level gradually increase above that of the reference case in the low frequency region, and then at $Re_c$ = 700,000 the flaplets begin to flutter in a similar manner to the long flaplets.
When observing the high frequencies, the reverse of the low frequency effect is observed, where the wide flaplets show the largest noise increase, followed by the baseline and the narrow flaplets. 
This is assumed to be due to the narrow flaplets being able to more successfully disrupt the small scale structures (due to their geometrical size) and vice versa for the low frequency reduction and the wide flaplets.  

At $\alpha_g = 10^\circ$, the low frequency reduction of the wide flaplets is comparable to that of the baseline flaplets, whereas the high frequency increase for the wide flaplets is still higher than that of the baseline flaplets. 
When tonal noise occurs, it can be seen that all the cases cause a delay and once again show a reduction. 
It is important to note that the narrow flaplets do not show any tonal noise across the whole $Re_c$ range. 
This is indicating that the narrow elements severely disrupt the acoustic feedback loop and the instabilities within the boundary layer on the pressure side of the aerofoil.
Increasing the angle to $\alpha_g = 15^\circ$ yields similar trends to those at lower angles. 
The exception is that the wide flaplets reach a higher Reynolds number before they go into flutter.  

\begin{figure}[t!] 
\centering\includegraphics[scale = 1]{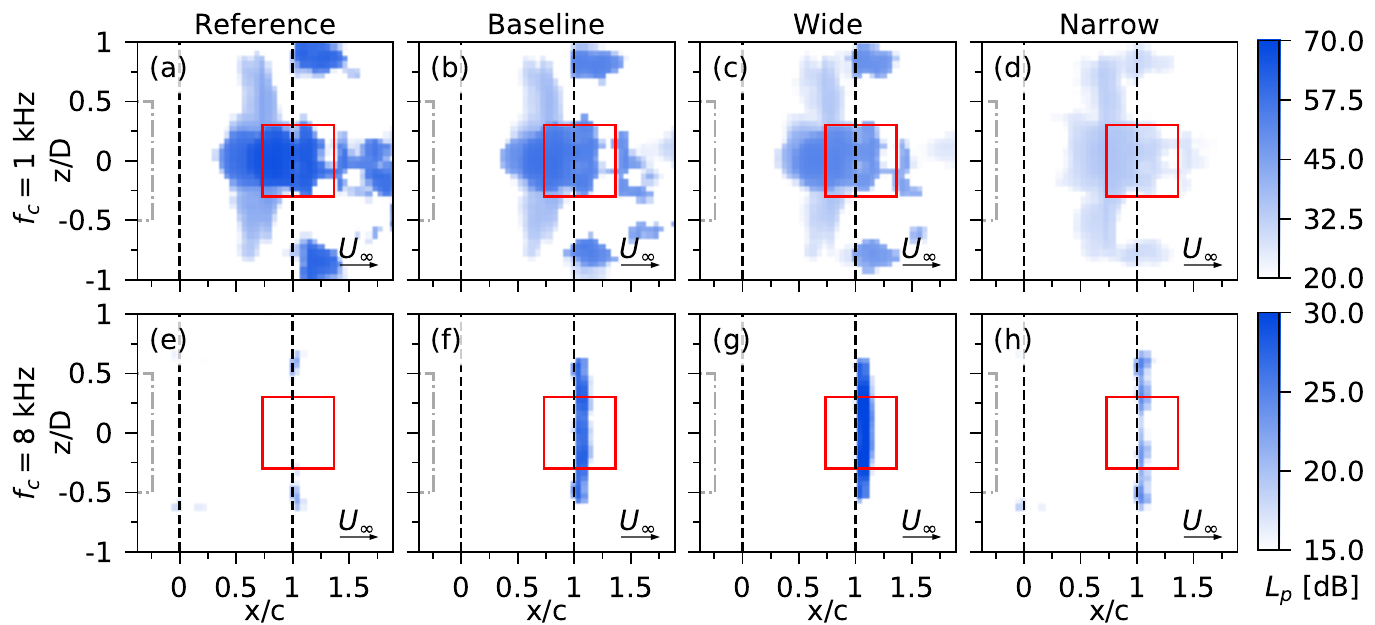}
\caption{2D sound maps as the width is varied, at $Re_c$=350,000 and $\alpha_g=10^\circ$. (a--d) shows the frequency band, $f_c$=1~kHz and (e--h) shows the frequency band, $f_c$ = 8~kHz. (\protect\begin{tikzpicture}
\protect\tikz[baseline=-2pt] \protect\draw[dash dot,very thick,color=lightgray] (0.0,0.1) -- (0.575,0.1);
\protect\end{tikzpicture}) indicates the jet nozzle, (\protect\begin{tikzpicture}
\protect\tikz[baseline=-2pt] \protect\draw[dashed,very thick,color=black] (0.0,0.1) -- (0.575,0.1);
\protect\end{tikzpicture}) indicates the aerofoil, (\protect\begin{tikzpicture}
\protect\tikz[baseline=-2pt] \protect\draw[solid,very thick,color=red] (0.0,0.1) -- (0.575,0.1);
\protect\end{tikzpicture}) indicates the interrogation region for acoustic spectra.}\label{fig: SoundMap-Width}
\end{figure} 

When looking at the sound maps for this geometric variation, Fig.~\ref{fig: SoundMap-Width}, at $f_c$=1~kHz the clear tonal noise reduction that was seen in Fig.~\ref{fig: AS-Width}b is seen with both the wide (c) and narrow (d) flaplets. 
The narrow flaplets show a significant reduction of up to 40~dB.
At $f_c$=8~khz, it can be clearly seen that the width of the flaplets has a strong effect on the high frequency acoustic scattering.
The wide flaplets (g) show a strong acoustic source across their surface, the strength of which decreases with reducing flaplet width, yielding a significantly lower acoustic source for the narrow flaplets (h) which is almost 15~dB lower than for the wide flaplets.
This again is showing that this high frequency noise is a function of the flaplet surface area.

\begin{figure}[t!] 
\centering\includegraphics[scale = 1]{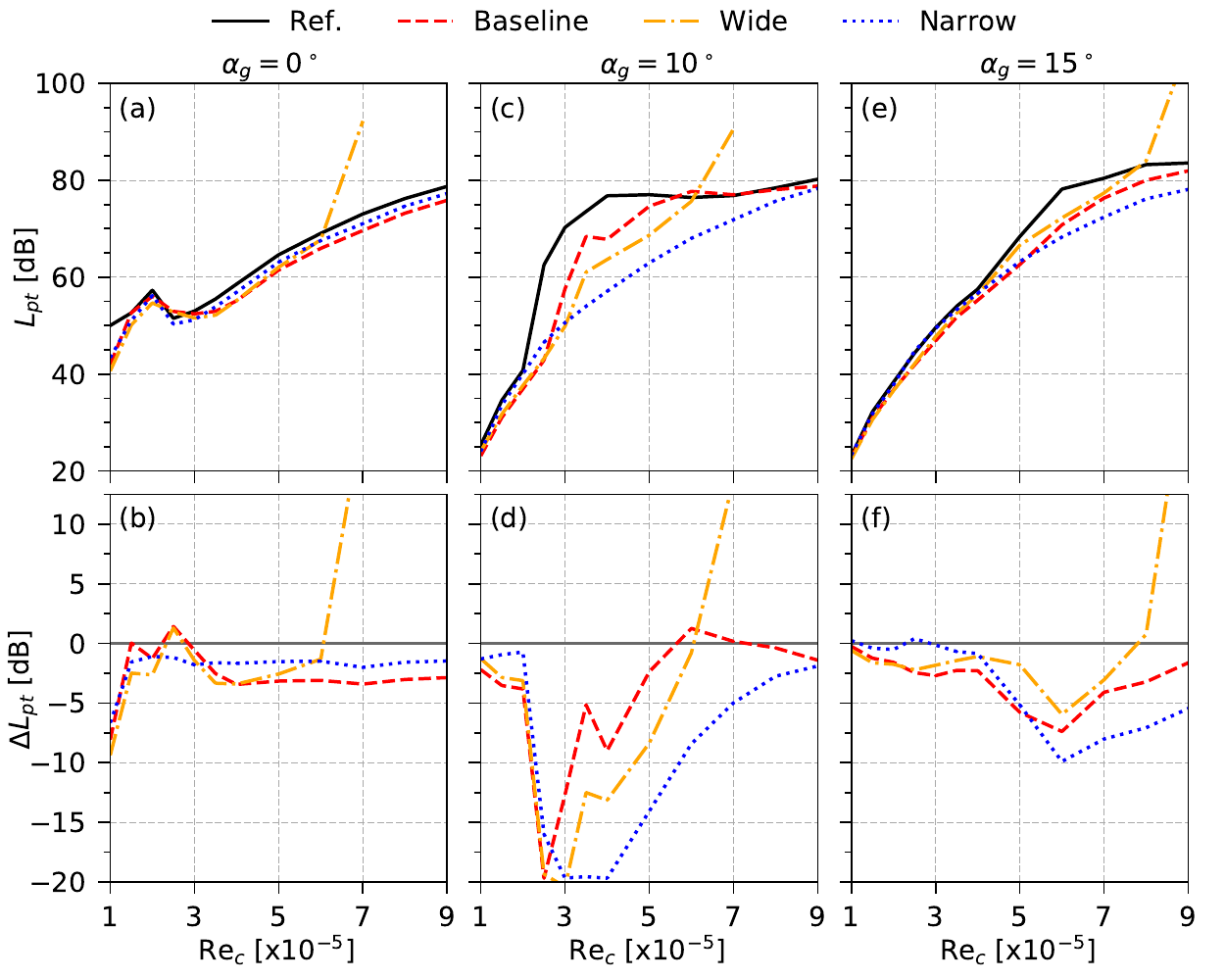}
\caption{OSPL, denoted as $L_{pt}$, and Delta OSPL, $\Delta L_{pt}$, for variation in width.}\label{fig: OSPL-Width}
\end{figure}

When looking at the $\Delta$OSPL at $\alpha_g = 0^\circ$, Fig.~\ref{fig: OSPL-Width}b, the narrow flaplets show a constant reduction of 2~dB across most of the tested velocity range.
The wide flaplets show a similar trend as the baseline flaplets, but at each Reynolds number below 400,000 there is more of a reduction. 
This corresponds to the greater low frequency reduction observed in Fig.~\ref{fig: AS-Width}a. 
Of course this trend ceases as the wide flaplets start to flutter. At $\alpha_g = 10^\circ$, it can be seen that for the low Reynolds number cases, up to $Re_c$ = 250,000, the narrow flaplets are not as efficient as the baseline and wide flaplets.
However after this point, when tonal noise is present, the narrow flaplets significantly out perform all other flaplet cases, with maximum reductions in the order of 20~dB observed.
As the tonal noise disappears for the reference aerofoil, the noise reduction tends towards 2.5~dB. 
Reductions of the OSPL can also be seen for the baseline and wide flaplets due to the slight suppression of the tonal noise component and low frequency reductions. 
As the angle increases further to $\alpha_g = 15^\circ$, reductions can be seen for all flaplets, and again the maximum reductions are visible when tonal noise is present on the reference aerofoil.

\begin{figure}[t!] 
\centering\includegraphics[scale = 1]{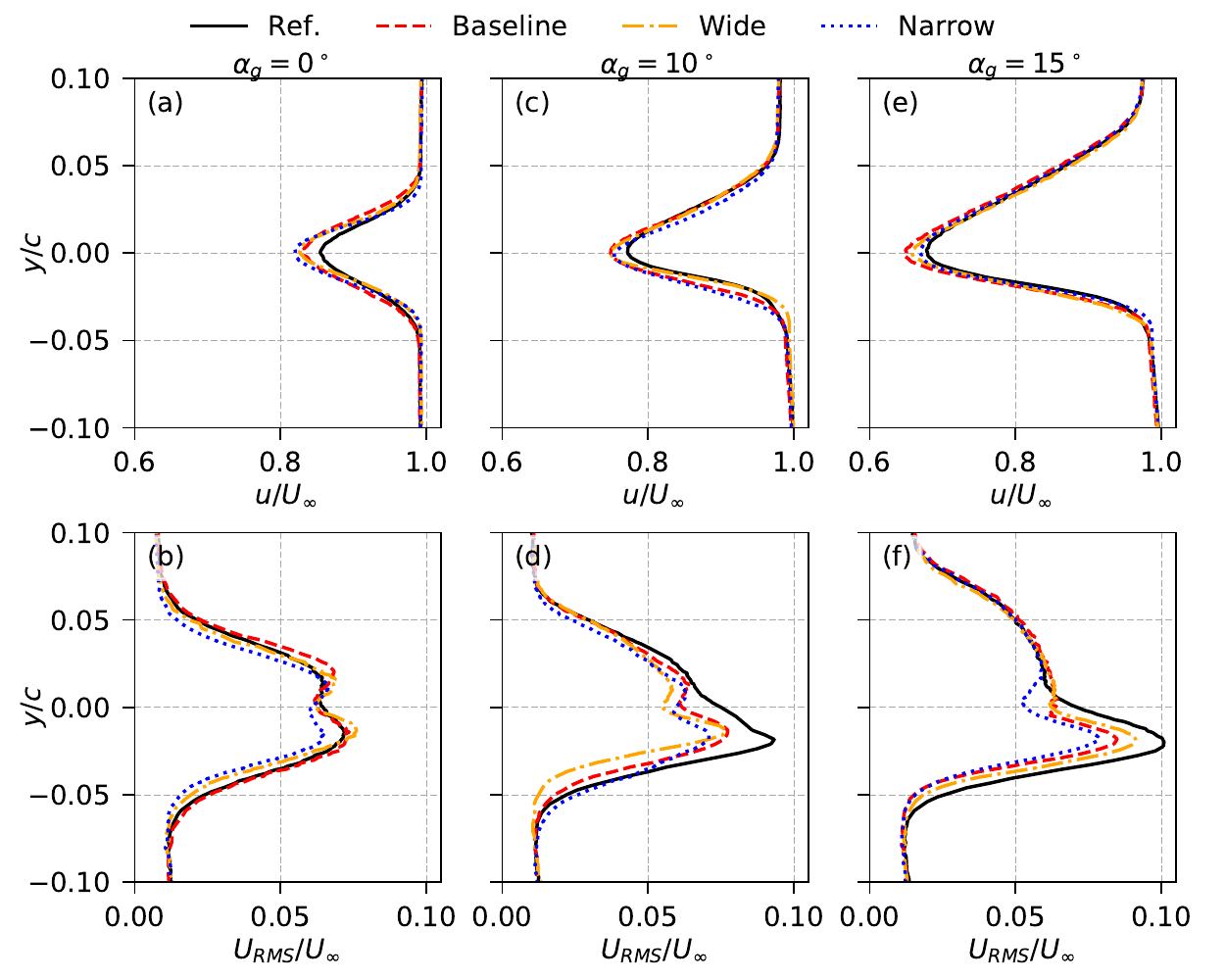}
\caption{Mean and RMS wake profiles of the streamwise velocity component at $Re_c = 200,000$ for variation in width.}\label{fig: HWA-Width}
\end{figure}

When looking at the hot wire measurements, the mean profiles in Figs.~\ref{fig: HWA-Width}a-c-e show that the flaplets all lead to a similar wake deficit.
Comparing the results with the wake deficit differences observed in Section~\ref{sec:Results sub:Length} for the flaplets of varying length leads to the conclusion that the differences there are only due to the geometrical variation of the length and not of the width. 
When looking at the RMS velocity profiles in Figs.~\ref{fig: HWA-Width}b-d-f, an interesting result can be seen in Fig.~\ref{fig: HWA-Width}f, where there is a clear order of the magnitude of the RMS velocity on the pressure side of the aerofoil.
It can be seen that the narrow flaplets dampen the turbulence in the wake most effectively, whilst the wide flaplets dampen the least.
Nevertheless, there is still a moderate reduction in comparison to the reference aerofoil.

\subsubsection{Variation in Flaplet Inter-Spacing}\label{sec:Results sub:Space}
\begin{figure}[t!] 
\centering\includegraphics[scale = 1]{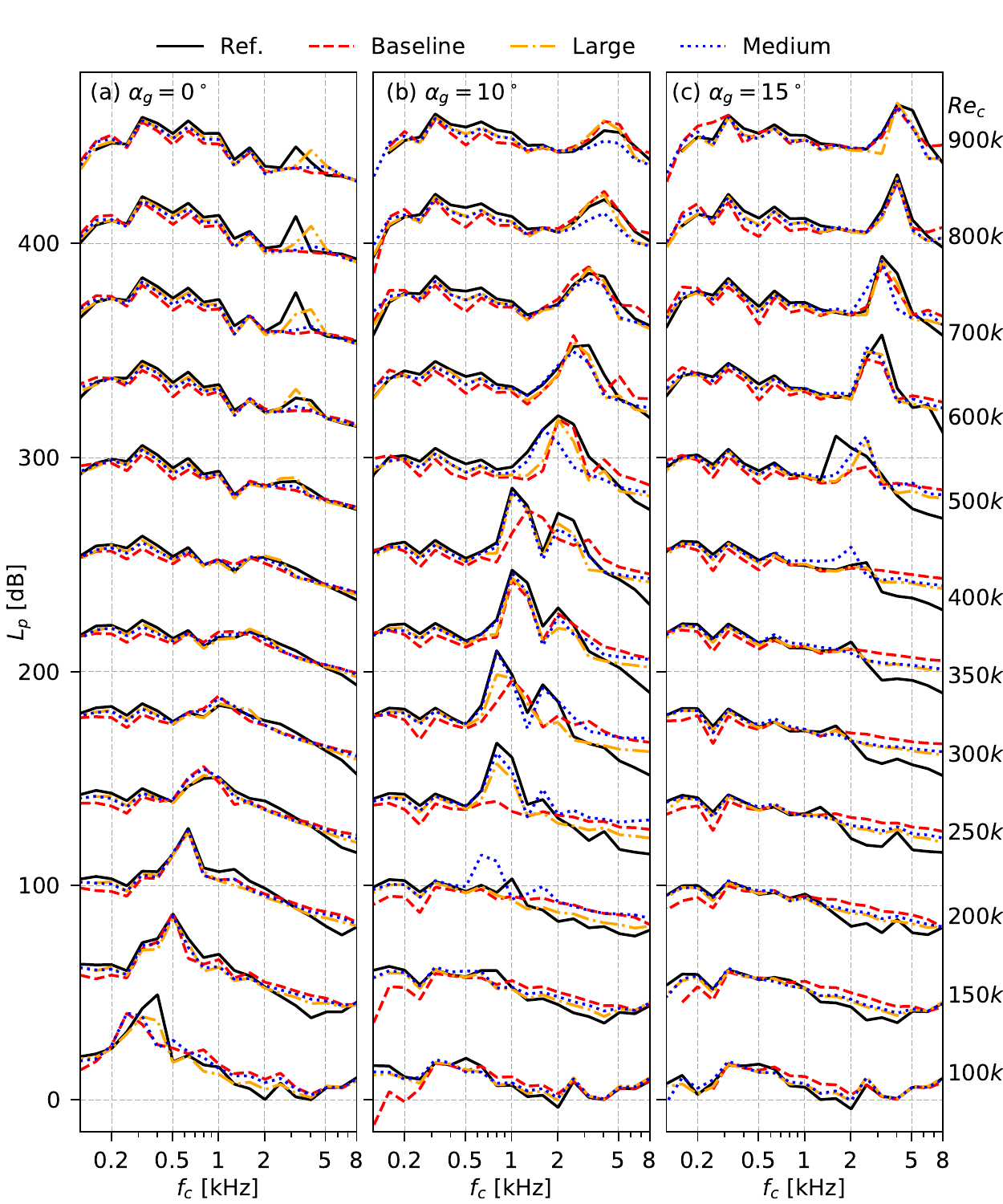}
\caption{1/3 Octave band acoustic spectra for variation in inter-spacing. Each of the spectra are spaced with 35~dB from each other for clarity.}\label{fig: AS-Space}
\end{figure}
The variation of the inter-spacing of the flaplets is in effect a way to alter the `porosity' of the flaplets.
The changes in the spacing are detailed in Table~\ref{table: Flaplets Spec}, and it can be seen that the smallest spacing is that of the baseline flaplets (1~mm) followed by the medium spacing (3~mm) and large spacing (7~mm).
The large spacing flaplets are spaced such that every other flaplet, from the baseline case, is removed. At $\alpha_g = 0^\circ$, Fig.~\ref{fig: AS-Space}a, the flaplets generally behave in a similar manner to each other. 
However, it is clear that the smallest spacing (baseline) has the most reduction in the low frequency range, whereas the medium and the large spaced flaplets show only a small or no reduction in this range. 
As is seen with all other tested cases, the flaplets do lead to an increase of noise in the high frequency range, the largest spacing inducing the smallest increase. 
When looking at the trailing-edge bluntness noise ($Re_c\,>$ 600,000 at $f_c = 3$~kHz), the baseline and medium flaplets show a reduction whereas the flaplets with large spacing show only a small reduction with regards to the baseline.
These results are explained by the fact that there is more trailing edge exposed, therefore the acoustic effect of the bluntness is more pronounced. 

Increasing the angle to $\alpha_g = 10^\circ$ and $15^\circ$ shows similar effects, and as in the previous geometric variations, the magnitude of the effects are enhanced.

\begin{figure}[t!] 
\centering\includegraphics[scale = 1]{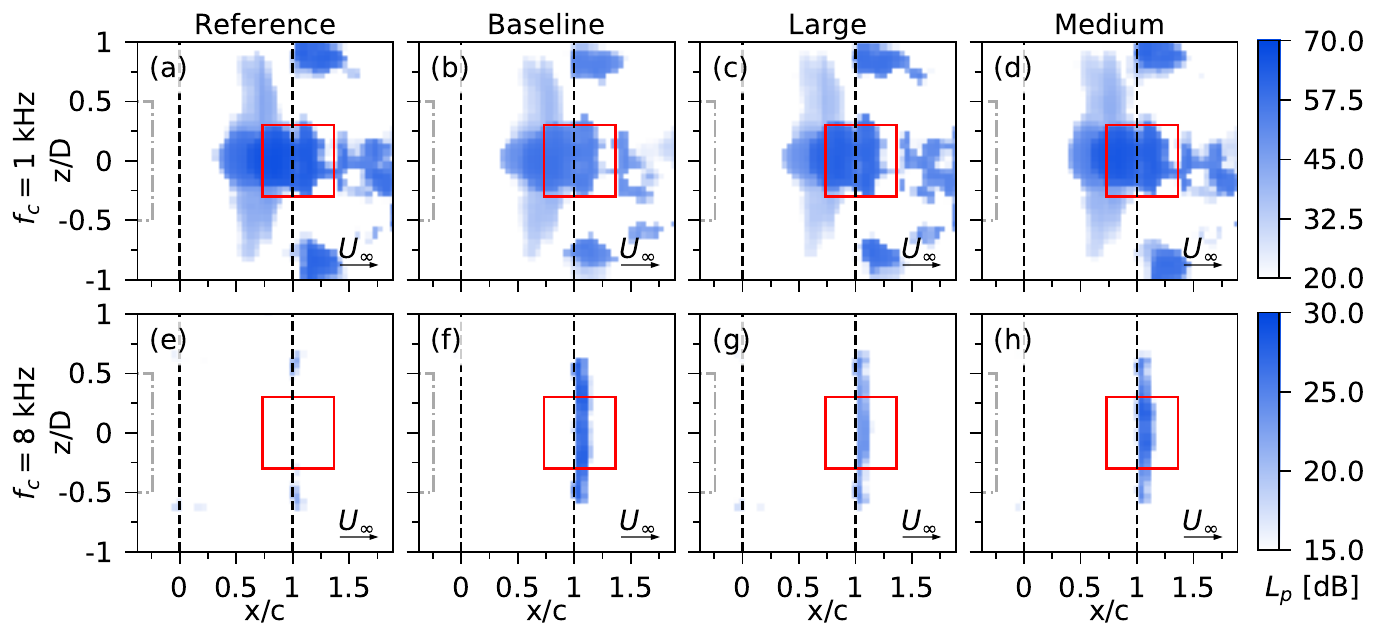}
\caption{2D sound maps as the inter-spacing is varied, at $Re_c$=350,000 and $\alpha_g=10^\circ$. (a--d) shows the frequency band, $f_c$=1~kHz and (e--h) shows the frequency band, $f_c$ = 8~kHz. (\protect\begin{tikzpicture}
\protect\tikz[baseline=-2pt] \protect\draw[dash dot,very thick,color=lightgray] (0.0,0.1) -- (0.575,0.1);
\protect\end{tikzpicture}) indicates the jet nozzle, (\protect\begin{tikzpicture}
\protect\tikz[baseline=-2pt] \protect\draw[dashed,very thick,color=black] (0.0,0.1) -- (0.575,0.1);
\protect\end{tikzpicture}) indicates the aerofoil, (\protect\begin{tikzpicture}
\protect\tikz[baseline=-2pt] \protect\draw[solid,very thick,color=red] (0.0,0.1) -- (0.575,0.1);
\protect\end{tikzpicture}) indicates the interrogation region for acoustic spectra.}\label{fig: SoundMap-Space}
\end{figure} 

The sound maps for the 1~kHz band shows very little difference between the different spacings.
At 8~kHz, it can be seen that as the spacing increases from the smallest (baseline case (f)) to the largest (g) the acoustic source located at the flaplets decreases in strength, again as with the other geometric variations this is due to the reduced flaplets surface area.

\begin{figure}[t!] 
\centering\includegraphics[scale = 1]{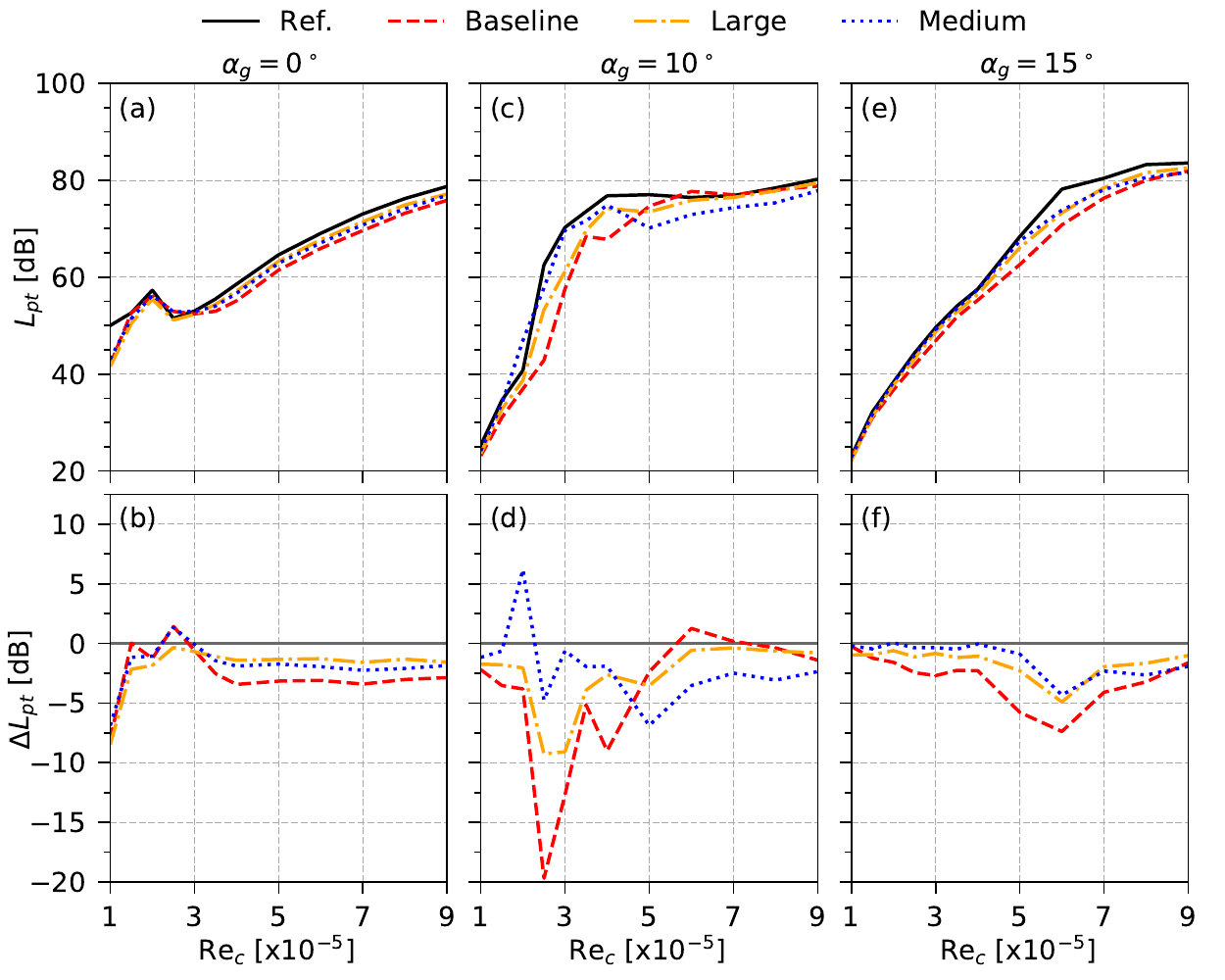}
\caption{OSPL, denoted as $L_{pt}$, and Delta OSPL, $\Delta L_{pt}$, for variation in inter-spacing.}\label{fig: OSPL-Space}
\end{figure}

When looking at the OSPL and the $\Delta$OSPL at $\alpha_g = 0^\circ$, Fig.~\ref{fig: OSPL-Space}a-b, it is visible that at flow velocities for which tonal noise ceases to occur ($Re_c>300,000$) there is a clear reduction of noise for all flaplets. Thereby, the magnitude of the reduction increases as the spacing is reduced.
A `steady' reduction of $\approx$ 3.5, 2.5 and 2~dB for each of the cases can be seen, in the order of inter-spacing distance.
This shows that, although the smallest spacing increases the high frequency noise the most, the low frequency noise reduction is sufficiently large enough to compensate and yield this moderate overall reduction.
At $\alpha_g = 10^\circ$, the tonal noise reduction is the dominant contribution to the overall noise reduction. 
Increasing the angle further shows an interesting result, where the medium spaced flaplets show no overall reduction until $Re_c=500,000$ where the reduction is again due to the suppression of the tonal noise component.
For this geometric constraint it can clearly be seen that, generally, the smallest spacing is the most efficient across all the tested cases. 

\begin{figure}[t!] 
\centering\includegraphics[scale = 1]{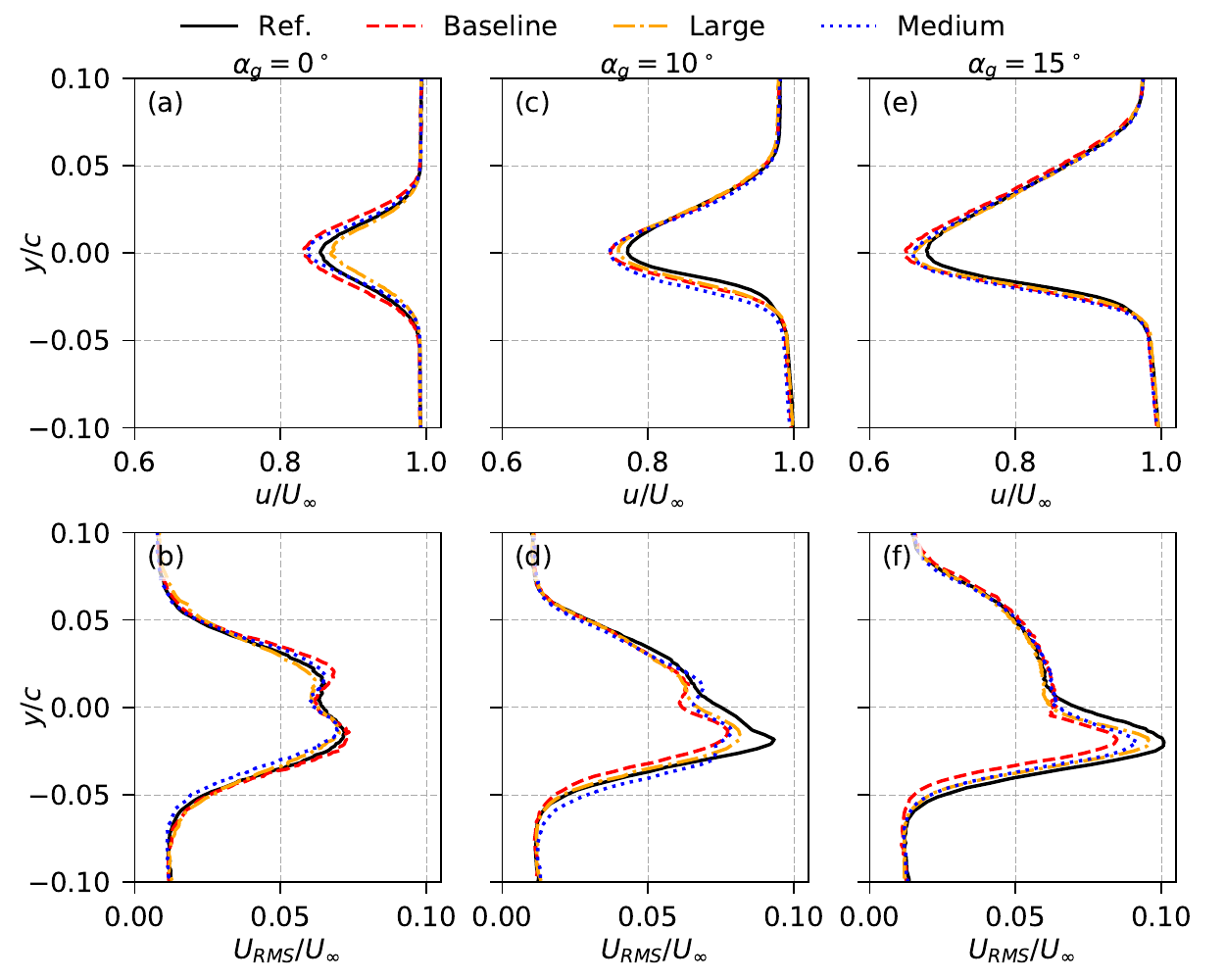}
\caption{Mean and RMS wake profiles of the streamwise velocity component at $Re_c = 200,000$ for variation in inter-spacing.}\label{fig: HWA-Space}
\end{figure}

Figure~\ref{fig: HWA-Space} shows the hot wire measurement results for this geometric variation. 
Again, as seen with the width variation, the difference observed is due to the elongation of the effective trailing edge. 
Looking at the RMS velocity profiles, again little difference is seen between inter-spacing at $\alpha_g = 0^\circ$ and $10^\circ$.
However, further increasing the angle to $15^\circ$ again shows a clear difference between the spacings, where the smallest spacing (baseline flaplets) leads to the smallest RMS values. 
As the spacing increases, the RMS velocity profile tends back towards that of the reference case, which is to be expected. 
Therefore it can be further seen that the most `optimal' configuration of spacing is indeed the shortest spacing. 

\subsection{Average $\Delta$OSPL}
\begin{table}[t]
\small
\centering
\begin{tabular}{p{0.22\textwidth}>{\centering}p{0.15\textwidth}>{\centering}p{0.15\textwidth}>{\centering\arraybackslash}p{0.15\textwidth}}
\hline
\multirow{2}{*}{Case} & \multicolumn{3}{c}{Average $\Delta L_{pt}$ [dB]}               \\
                      & $\alpha_g=0^\circ$ & $\alpha_g=10^\circ$ & $\alpha_g=15^\circ$ \\ \hline
Baseline              & -2.5               & -4.9                & -2.9                \\
Long                  & -1.0               & -0.7                & -0.2                \\
Short                 & -3.2               & -8.0                & -3.8                \\
Wide                  & -0.6               & -6.8                & 0.3                \\
Narrow                & -2.0               & -9.2                & -3.1                \\
Large Spacing           & -1.9               & -3.0                & -1.6                \\
Medium Spacing            & -1.8               & -1.9                & -1.2                \\ \hline
\end{tabular}
\caption{Average overall noise difference, in comparison to the reference aerofoil.}\label{table: AvgOSPL}
\end{table}

In order to obtain an overview of the overall effect of the different flaplets, the average of the $\Delta$OSPL for each Reynolds number case and angle of attack has been collected in Table~\ref{table: AvgOSPL}, helping to indicate which configuration is generally considered to be the best option.
At $\alpha_g = 0^\circ$, it can be seen that the short flaplets are the best option. 
This is primarily due to the reduction in the low to mid frequency range, which also led to the reduction of the tonal peak. 
All other cases, apart from the long and wide flaplets, also show a reasonable reduction at this test case. 
Increasing to $\alpha_g = 10^\circ$ shows the largest overall reduction for the narrow flaplets with an average reduction of 9.2~dB. 
This is followed by the short flaplets which yield an average reduction of 8.1~dB. 
These reductions are due to the almost complete suppression of the tonal noise component for these cases. 
Both the narrow flaplets and the short flaplets also have the additional benefit of generating only a reduced amount of high frequency noise in comparison to the other tested cases. 
Hence, both of these benefits add to give significant reductions. 
Finally, at $\alpha_g = 15^\circ$, the short and the narrow flaplets once again show the best performance, which is again due to the reduction of tonal noise at the highest Reynolds numbers. 

\section{Conclusion}
Herein, an extensive acoustic study has been carried out on the geometric optimisation of self-oscillating trailing edge flaplets to reduce aerofoil self-noise in a passive way.
The effect of a variation of the length, width and inter-spacing of these flaplets has been investigated against the reference case of a plain aerofoil.
Basically, it has been shown that all flaplets show a reduction in tonal noise, to some extent. 
This has been attributed to the flaplets working as a dampener on T-S instabilities in the boundary layer. 
The most effective tonal noise reduction was seen with the narrow flaplets, where almost all tonal components were dampened out. 
Besides the narrow flaplets, the short flaplets were also extremely effective at $Re_c > 600,000$. 

As has been seen already in previous studies on trailing edge flaplets by the authors \citep{Talboys2019, Talboys2019a}, the flaplets with the small inter-spacing show a clear low frequency noise reduction.
Furthermore, the range of frequencies where the most reduced noise level is achieved can be tuned by the  Eigen frequency of the flaplets, herein by altering the length of the flaplets.
When the long flaplets (having the lowest Eigen frequency) were tested, the noise level was reduced at a lower frequency range and vice versa for the short flaplets (which are characterised by a higher Eigen frequency).
This observation clearly supports the hypothesis of a lock-in mechanism, as introduced by \citet{Talboys2018} as an underlying physical principle that explains the dampening of the non-linear growth of instabilities in the boundary layer along the aerofoil surface. 
According to this, the flaplets are considered as flat cantilever beams, which are excited by the flow to oscillate into their fundamental bending mode at the fundamental Eigen frequency. 
These oscillations, when close to the natural frequency of the instabilities in the boundary layer, freeze the instabilities in their linear state, which is called a lock-in \citep{Talboys2018}. 
Therefore, one can consider the flaplets as effective pacemakers to the flow instabilities.    

Concerning the width of the flaplets, it was found that the widest flaplets had the most effect within the low frequency range and the narrow flaplets were least effective, albeit a slight reduction was still observed.
When looking at the high frequency noise increase, it was most evident that the narrow flaplets show little noise increase, especially in comparison with the wide flaplets. 
It can be assumed that this is due to the loss of spanwise coherence for the narrow flaplets, which are free to move independently from each other.

An interesting feature was observed at high Reynolds numbers ($Re_c>600,000$), where two of the geometries show a sudden massive increase in noise emission. 
This is attributed to the flaplets going past their critical velocity, at which the torsional bending mode is excited. This causes flow separation due to flutter and the flaps are no longer  effective. 
Therefore, this constraint must be accounted for in future designs.
This could be done either by modifying the flaplets geometry or via the use of non-isotropic materials to prevent the torsional mode.

When looking at the flow profiles in the wake, it can be seen that the acoustic results are correlated to the peak levels of turbulence (RMS values of the velocity fluctuation) in the wake, measured by the hot-wire anemometer. 
All flaplets did show some level of reduction of the turbulence. The lowest turbulence levels were observed for the narrow flaplets, showing that this geometry in particular is good at disrupting the turbulence convected downstream of the trailing edge.

Following this extensive study, it can be concluded that an optimal flaplet geometry might have a combination of both short and narrow flaplets to see if the benefits of both modifications will hold together and produce an even more effective flaplet configuration. 
This is left as future work. 

\section*{Acknowledgements}
The position of Professor Christoph Br\"{u}cker is co-funded by BAE Systems and the Royal Academy of Engineering (Research Chair no. RCSRF1617$\backslash$4$\backslash$11) and funding for Mr. E. Talboys to carry out the experiments in Cottbus was made available by The Worshipful Company of Scientific Instrument Makers (WCSIM), both of which are gratefully acknowledged.





\bibliographystyle{model1-num-names}
\bibliography{TrailingEdgeFlaps.bib}







\end{document}

%% file: Fig_Tex/MicArray.tex
\begin{figure}[t!]
\begin{minipage}[b]{0.48 \linewidth}
\centering 
\subfloat[Schematic display of the measurement setup, plan view]{\label{fig: Exp Arr. Acc}
\small
\begin{tikzpicture}[scale=3.5,>=latex]
		\draw[thick,blue,fill,fill opacity=0.2] (-0.48,-0.14) rectangle (-0.28,0.14);
		\draw[blue](-0.3,-0.14) -- +(-45:0.3) node [below] {aerofoil at $z=0$~m};
		\begin{scope}[scale=0.2,xshift=-2.65cm]
			\draw[very thick,yellow](-2,2) node [above right,text width=2cm] {nozzle} .. controls (-1,2) and (-0.7,0.5) .. (0,0.5);
			\draw[very thick,yellow](-2,-2) .. controls (-1,-2) and (-0.7,-0.5) .. (0,-0.5);
			\shadedraw[dotted,left color=gray!40,opacity=0.8](5,0) -- (0,0.5) -- (5,1.5);
			\shadedraw[dotted,left color=gray!40,opacity=0.8](5,0) -- (0,-0.5) -- (5,-1.5);
			\draw[->](-1,0.25)--(0,0.25);
			\draw[->](-1,0)--(0,0) node [at start,left]{};
			\draw[->](-1,-0.25)--(0,-0.25);
			\draw(2.2,0) -- +(122.5:3.2) node [above left] {core jet};
			\draw(2.2,0.7) -- +(110:3.2) node [above] {mixing zone};
		\end{scope}
		\draw[red,only marks,mark=*,mark options={scale=0.15}] plot coordinates {
		(-0.145527,0.6335)
		(-0.066774,0.236764)
		(-0.376879,0.529587)
		(-0.152297,0.193188)
		(-0.550855,0.345049)
		(-0.214634,0.120201)
		(-0.640968,0.10798)
		(-0.244295,0.028914)
		(-0.6335,-0.145527)
		(-0.236764,-0.066774)
		(-0.529587,-0.376879)
		(-0.193188,-0.152297)
		(-0.345049,-0.550855)
		(-0.120201,-0.214634)
		(-0.10798,-0.640968)
		(-0.028914,-0.244295)
		(0.145527,-0.6335)
		(0.066774,-0.236764)
		(0.376879,-0.529587)
		(0.152297,-0.193188)
		(0.550855,-0.345049)
		(0.214634,-0.120201)
		(0.640968,-0.10798)
		(0.244295,-0.028914)
		(0.6335,0.145527)
		(0.236764,0.066774)
		(0.529587,0.376879)
		(0.193188,0.152297)
		(0.345049,0.550855)
		(0.120201,0.214634)
		(0.10798,0.640968)
		(0.028914,0.244295)
		(-0.029598,0.463341)
		(-0.02879,0.088281)
		(-0.348561,0.306703)
		(-0.082782,0.042067)
		(-0.463341,-0.029598)
		(-0.088281,-0.02879)
		(-0.306703,-0.348561)
		(-0.042067,-0.082782)
		(0.029598,-0.463341)
		(0.02879,-0.088281)
		(0.348561,-0.306703)
		(0.082782,-0.042067)
		(0.463341,0.029598)
		(0.088281,0.02879)
		(0.306703,0.348561)
		(0.042067,0.082782)
		(-0.138631,0.241627)
		(-0.268883,0.072829)
		(-0.241627,-0.138631)
		(-0.072829,-0.268883)
		(0.138631,-0.241627)
		(0.268883,-0.072829)
		(0.241627,0.138631)
		(0.072829,0.268883)
		};
	    \draw[red](0.54,0.6) node [above,text width=2.25cm] {microphones ($z=0.7$~m)};
		\draw[->] (-0.932,-0.72) -- (-0.932,0.8) node [left] {$y$ [m]};
		\draw[->] (-0.932,-0.72) -- (0.75,-0.72);
		\draw[black](0, -0.875) node [below]{$x$ [m]};
		\foreach \x in {-0.53,-0.28,0,0.5}
			 \draw (\x,-0.7) -- +(0,-0.04) node [below] {\x};
		\foreach \y in {-0.2,-0.1,0,0.1,0.2} 
			 \draw (-0.912,\y) -- +(-0.04,0) node [left] {\y};		
	\end{tikzpicture}}\normalsize
\end{minipage}
\begin{minipage}[b]{0.67\linewidth}
\centering
\subfloat[Photograph of experimental set-up, looking from downstream of the aerofoil and nozzle.\label{fig: Aerofoil Pic}]{\includegraphics[scale=0.65]{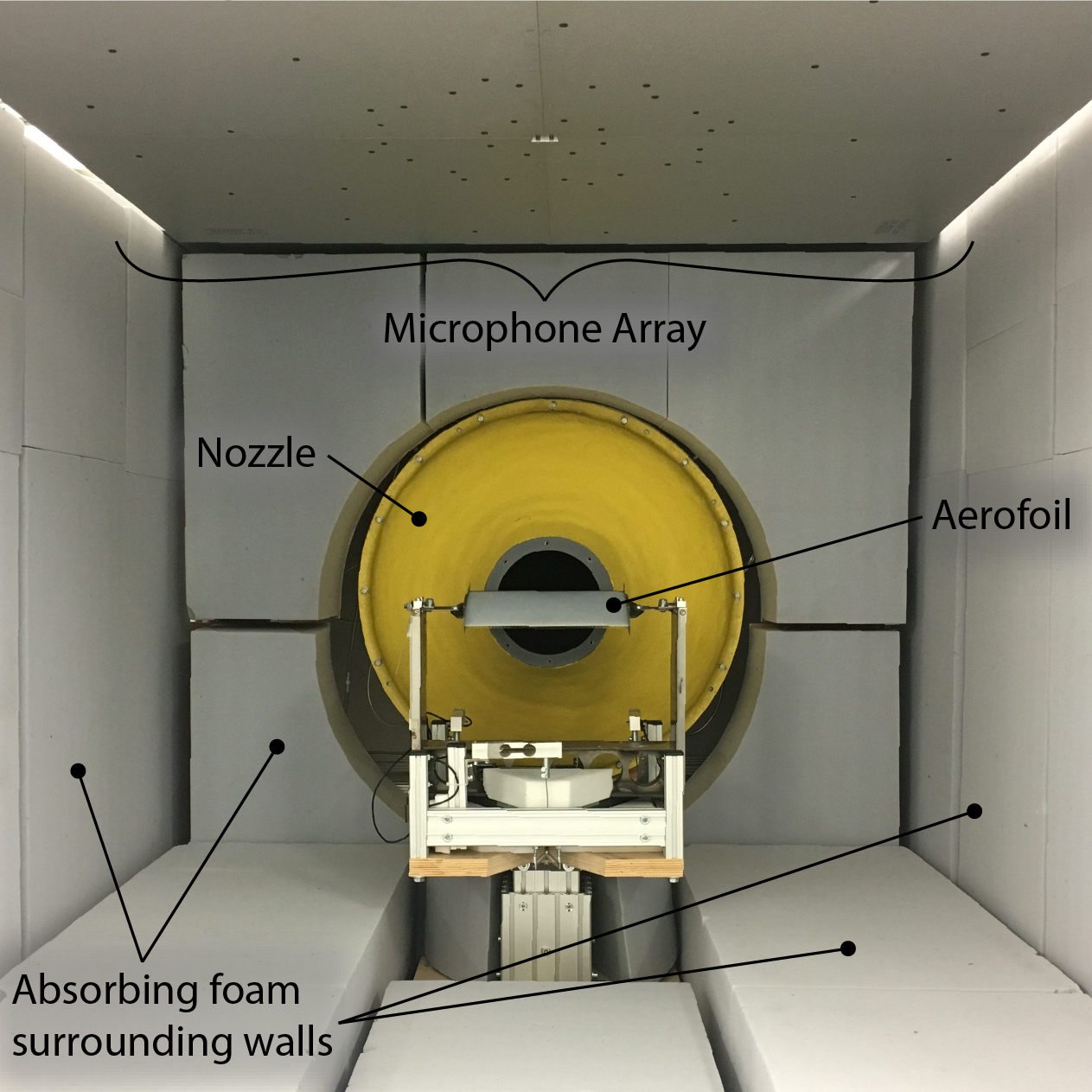}}
\end{minipage}
\medskip\par
\begin{minipage}[b]{\linewidth}	
\centering
\subfloat[Sketch of the aerofoil with the flaplets attached to the trailing edge and the naming convention for the geometric properties of the flaplets.\label{fig: Aerofoil Pic}]{\includegraphics[scale=1.25]{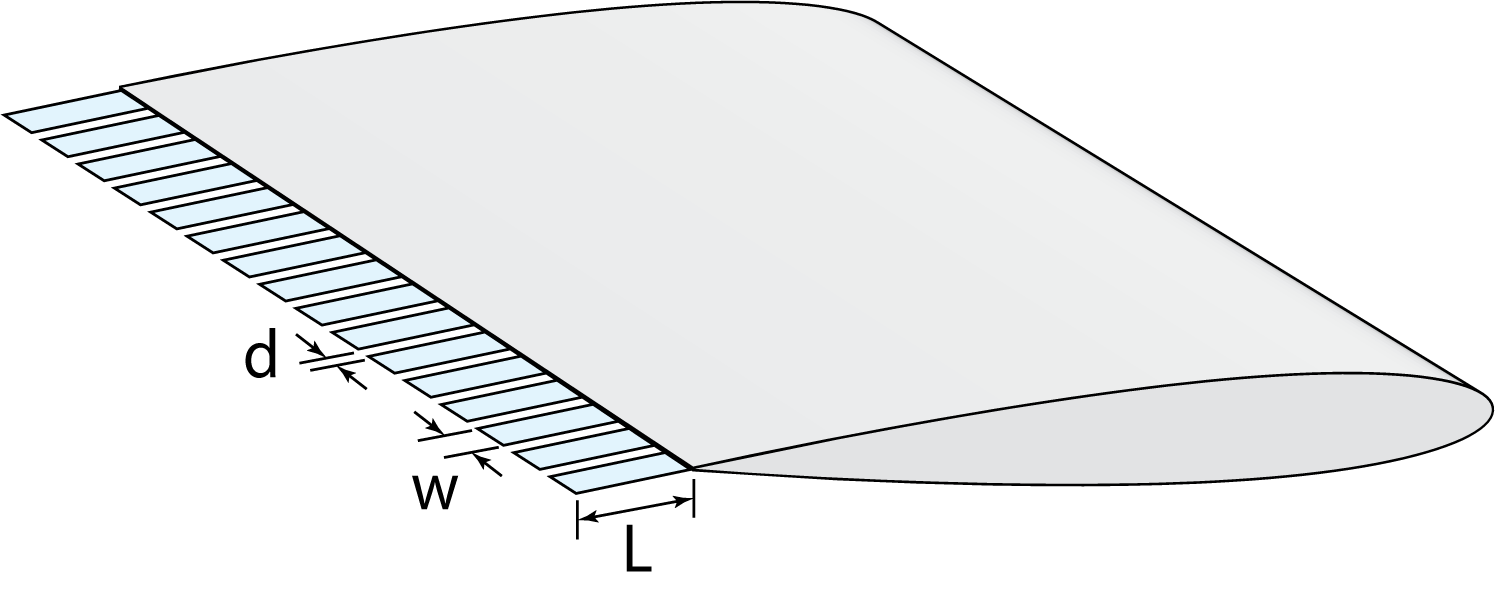}}
\end{minipage}
\caption{Overview of experimental set-up.}
	\label{fig: Set-up}
\end{figure}